\documentclass[fleqn]{elsart1p}

\usepackage{amsmath}
\usepackage{tabularx}
\usepackage{rotating}
\usepackage{float}
\usepackage{setspace}
\usepackage{graphics}
\usepackage{graphicx}
\usepackage{amssymb}

\makeatletter
\newcommand{\sech}{\mathop{\operator@font sech}}
\makeatother

\newcommand{\nablae}{ \widetilde{\nabla}}

\usepackage{graphics}
\usepackage{graphicx}
\usepackage{amssymb}

\begin{document}

\begin{frontmatter}

\title{Boussinesq systems in two space dimensions
over a variable bottom for the generation and propagation of
tsunami waves\thanksref{2}}
\thanks[2]{Research supported by the Sixth Framework Programme of EU project TRANSFER (Tsunami Risk ANd Strategies For the European Region) under contract no. 037058.}

\author{D. E. Mitsotakis}
\ead{dmitsot@math.uoa.gr}
\address{Institute of Applied and Computational Mathematics, FO.R.T.H., P.O. Box 1527,
71110 Heraklion, Greece}

\begin{abstract}
Considered here are Boussinesq systems of equations of surface
water wave theory over a variable bottom. A simplified such Boussinesq system is derived and solved numerically by the standard Galerkin-finite element method. We study by numerical means the generation of tsunami waves due to bottom deformation and we compare the results with analytical solutions of the linearized Euler equations. Moreover, we study tsunami wave propagation in the case of the Java 2006 event, comparing the results of the Boussinesq model with those produced by the finite difference code MOST, that solves the shallow water wave equations. 
\end{abstract}
\end{frontmatter}
% main text

\section{Introduction}

In recent years, there have been many theoretical and computational advances in the study of the full water-wave problem. However, there is still need for accurate, simpler mathematical models. Boussinesq systems are systems of partial differential equations that approximate the three-dimensional Euler equations that describe the irrotational, free surface flow of an incompressible, inviscid fluid. Because of their simplicity, Boussinesq systems have been used in the study of a variety of water wave phenomena in ports, channels, coastal areas, and in the open sea. They have also been used in studies of tsunami wave generation and propagation, cf., e.g.,  \cite{GHM}, \cite{LBLS}, \cite{Wu}.

In the case of horizontal bottom, these systems, derived in their general form in \cite{BCS1}, \cite{BCL}, may be written in nondimensional, unscaled variables as 
\begin{equation}\label{E1.1}
\begin{array}{c}
\eta_t+\nabla\cdot {\bf u}+\nabla\cdot \eta{\bf
u}+a\Delta\nabla\cdot {\bf u} -b\Delta\eta_t=0,\\
{\bf u}_t+\nabla\eta+\frac{1}{2}\nabla|{\bf
u}|^2+c\Delta\nabla\eta-d\Delta{\bf u}_t=0.
\end{array}
\end{equation}
In these equations, the independent variable ${\bf x}=(x,y)$
represents the position, $t$ is proportional to elapsed time,
$\eta=\eta({\bf x},t)$ is proportional to the deviation of the
free surface from its rest position, while ${\bf u}={\bf u}({\bf
x},t)=(u_1({\bf x},t),u_2({\bf x},t))^T$ is proportional to the
horizontal velocity of the fluid at some height. The coefficients
$a,b,c,d$ are given by the formulas
\begin{equation}\label{E1.2}
a=\frac{1}{2}(\theta^2-\frac{1}{3})\nu,\,\,b=\frac{1}{2}(\theta^2-\frac{1}{3})(1-\nu),\,\,
c=\frac{1}{2}(1-\theta^2)\mu,\,\,d=\frac{1}{2}(1-\theta^2)(1-\mu),
\end{equation}
where $\nu,\mu$ are real constants and $0\le\theta\le 1$. (If we
denote the nondimensional depth variable by $z$ (with positive direction upwards), then the bottom
of the channel lies at $z=-1$, while the horizontal velocity ${\bf u}$ 
is evaluated at the nondimensional height $z=-1+\theta (1+\eta({\bf x},t))$.)

As it is explained in detail in \cite{BCS1}, the Boussinesq
approximation on which (\ref{E1.1}) is based is valid when
$\varepsilon:=A/h_0\ll 1$, $\sigma:=\lambda/h_0\gg 1$, and the Stokes number
$S=A\lambda^2/h_0^3$ is of order 1; here $A$ is the maximum free
elevation above the undisturbed level of the fluid of depth $h_0$
and $\lambda$ a typical wavelength. Letting $S=1$ and working in
scaled, nondimensional variables, one may derive from the Euler
equations, by appropriate expansion in powers of $\varepsilon$,
the scaled version of (\ref{E1.1}) in the form
\begin{equation}\label{E1.3}
\begin{array}{c}
\eta_t+\nabla\cdot {\bf u}+\varepsilon\left[\nabla\cdot \eta{\bf
u}+a\Delta\nabla\cdot {\bf u} -b\Delta\eta_t\right]=O(\varepsilon^2),\\
{\bf u}_t+\nabla\eta+\varepsilon\left[\frac{1}{2}\nabla|{\bf
u}|^2+c\Delta\nabla\eta-d\Delta{\bf u}_t\right]=
O(\varepsilon^2),
\end{array}
\end{equation}
from which (\ref{E1.1}) follows by rescaling and replacing the
right-hand side by zero.

It is worthwhile to mention that in \cite{BCL} a new family of Boussinesq systems was derived. These are the {\em fully symmetric} Boussinesq systems of general form
\begin{equation}\label{E1.4}
\begin{array}{l}
\eta_t+\nabla\cdot {\bf u}+\frac{1}{2}\nabla\cdot
\eta{\bf
u}+a\Delta\nabla\cdot {\bf u} -b\Delta\eta_t=0,\\
{\bf u}_t+\nabla\eta+\frac{1}{4}\nabla\eta^2+\frac{3}{2}\begin{pmatrix}\partial_x u_1^2\\
\partial_y u_2^2
\end{pmatrix}+\frac{1}{4}\begin{pmatrix}\partial_x u_2^2\\
\partial_y u_1^2
\end{pmatrix}+\frac{1}{2}\begin{pmatrix}\partial_y (u_1 u_2)\\
\partial_x (u_1u_2)
\end{pmatrix}+
c\Delta\nabla\eta-d\Delta{\bf u}_t=0,
\end{array}
\end{equation}
where $a,b,c,d$ are given by (\ref{E1.2}). These systems, like the usual Boussinesq systems (\ref{E1.1}), are formally $O(\varepsilon^2)$ approximations of the Euler equations when written in nondimensional, scaled variables in a form similar to that of (\ref{E1.3}), wherein their nonlinear and dispersive terms are multiplied by $\varepsilon$.

In addition, several types of Boussinesq systems with variable bottom have been derived, cf. e.g., \cite{Chazel}, \cite{C3}, \cite{DD}, \cite{MMS}, \cite{MS}, \cite{N}, \cite{P}, \cite{SM}. The study of Boussinesq systems with variable bottom was initiated by Peregrine, \cite{P}, who derived the system
\begin{equation}\label{E1.5}
\begin{array}{l}
\eta_t+\nabla\cdot [(h+\varepsilon \eta)\bar{{\bf u}}]=0,\\
\bar{{\bf u}}_t+\nabla\eta+\varepsilon(\bar{{\bf
u}}\cdot\nabla)\bar{{\bf u}}
-\sigma^2\frac{h}{2}\nabla(\nabla\cdot (h\bar{{\bf
u}}_t))+\sigma^2\frac{h^2}{6}\nabla(\nabla\cdot \bar{{\bf
u}}_t)=O(\varepsilon
\sigma^2,\sigma^4),
\end{array}
\end{equation}
where $\bar{{\bf u}}$ denotes the depth-averaged velocity, i.e.
\begin{equation*}
\bar{{\bf u}}=\frac{1}{h+\varepsilon\eta}
\int_{-h}^{\varepsilon\eta}{\bf u}dz.
\end{equation*}
Using (\ref{E1.5}) one may derive other Boussinesq systems. We mention, for example, Nwogu's system, \cite{N}, 
\begin{equation}\label{E1.6}
\begin{array}{l}
\eta_t+\nabla\cdot((\varepsilon\eta+h){\bf
u})+\sigma^2\nabla\cdot\left[
\left(\theta-\frac{1}{2}\right)h^2\nabla(\nabla\cdot(h{\bf
u}))+\left(\frac{\theta^2}{2}-\theta+\frac{1}{3}\right)
h^3\nabla(\nabla\cdot {\bf u})\right]=O(\varepsilon\sigma^2,\sigma^4),\\
{\bf u}_t+\nabla\eta+\varepsilon({\bf u}\cdot
\nabla){\bf
u}+\sigma^2\left[(\theta-1)h\nabla(\nabla\cdot(h{\bf
u}_t)) +(\theta-1)^2\frac{h^2}{2}\nabla(\nabla\cdot {\bf
u}_t)
\right]=O(\varepsilon\sigma^2,\sigma^4),
\end{array}
\end{equation}
where ${\bf u}$ denotes the horizontal velocity as in the case of the systems (\ref{E1.1}).

In the paper at hand, we extend the Boussinesq models (\ref{E1.5}) and (\ref{E1.6}) in the case of
a bottom that depends on $x$, $y$ and $t$ to classes of systems analogous to (\ref{E1.1}) and (\ref{E1.4}). We also derive a simplified version of a BBM-type system appropriate for solving numerically realistic wave propagation problems by the finite element method. Theoretical and numerical aspects of these systems in the case of a horizontal bottom, i.e., for the systems (\ref{E1.1}) and (\ref{E1.4}), were studied recently in \cite{BCL}, \cite{C2}, \cite{DMS1}, \cite{DMS2}. 

We then apply the standard Galerkin-finite element method to the simplified Boussinesq model with homogeneous Dirichlet boundary conditions, and study the generation and propagation of tsunami waves. We compare the tsunami waves generated by this Boussinesq model and by the linearized Euler equations. Finally we study the propagation of a tsunami wave for a real tsunami event (that affected the island of Java in July 17, 2006), comparing the Boussinesq and the MOST models, \cite{TS}. The MOST model is an efficient numerical code solving the shallow water wave equations combining a finite difference scheme and a splitting technique, cf. \cite{TS}. We do not study here the inundation caused by the tsunami, as the Boussinesq code is not yet equipped with a runup algorithm.

\section{Boussinesq systems over variable bottom}\label{bsvb}

\subsection{Generalization of (\ref{E1.1})}

We denote by $(\tilde{x},\tilde{y},\tilde{z})$ a Cartesian
coordinate system. with $\tilde{z}$ measured upwards from the still water
level. Consider a three-dimensional wave field with water-surface
deviation propagating from its rest position,
$\tilde{\eta}(\tilde{x},\tilde{y},\tilde{t})$, at time
$\tilde{t}$, over a variable bottom given by $\tilde{h}(\tilde{x},\tilde{y},\tilde{t})=
\tilde{D}(\tilde{x},\tilde{y})+\tilde{\zeta}(\tilde{x},\tilde{y},\tilde{t})$. (We will assume that the variation of the time-dependent part of the bottom $\tilde{\zeta}$ is of the same order of magnitude as the surface elevation $\tilde{\eta}$.)
The fluid velocity is denoted
by $\hat{{\bf u}}=(\tilde{u},\tilde{v},\tilde{w})^T$.
The Euler equations, which describe three dimensional wave
propagation on the free surface, \cite{W}, are written in the form
\begin{eqnarray}
\hat{{\bf u}}_{\tilde{t}}+(\hat{{\bf u}}\cdot \nablae)\hat{{\bf
u}}+\frac{1}{\rho}\nablae \tilde{P}= -g{\bf k}, & & \label{E2.1}\\
\nablae\cdot \hat{{\bf u}}=0,& &\,\,\,\mbox{for}\,\,-\tilde{h}<
\tilde{z}<\tilde{\eta}(\tilde{x},\tilde{y},\tilde{t}),\label{E2.2}\\
\nablae\times \hat{{\bf u}}=0, & & \label{E2.3}
\end{eqnarray}
where $\tilde{P}$ is the pressure field, $\rho$ is the density,
$g$ is the acceleration due to gravity, ${\bf k}=(0,0,1)^T$ and
$\nablae=(\partial_{\tilde{x}},\partial_{\tilde{y}},\partial_{\tilde{z}})^T$.
The first equation expresses the conservation of momentum, whereas
the other two equations express the conservation of mass and the
irrotationality of the flow, respectively. The kinematic boundary
conditions at the free surface and bottom can be expressed as
\begin{equation}\label{E2.4}
\tilde{\eta}_{\tilde{t}}+\hat{{\bf u}}\cdot
\nablae\tilde{\eta}=0,\,\,\,\mbox{for}\,\,\tilde{z}=\tilde{\eta}(\tilde{x},\tilde{y},\tilde{t}),
\end{equation}
and
\begin{equation}\label{E2.5}
\tilde{h}_{\tilde{t}}+\hat {{\bf u}}\cdot\nablae
\tilde{h}=0,\,\,\,\mbox{for}\,\,\tilde{z}=-\tilde{h}(\tilde{x},\tilde{y},\tilde{t}),
\end{equation}
respectively. The fluid is assumed to satisfy the dynamic boundary
condition $\tilde{P}=0$ at the free
surface $\tilde{z}=\tilde{\eta}(\tilde{x},\tilde{y},\tilde{t})$.

Consider a characteristic water depth $h_0$, a typical wavelength
$\lambda_0$ and a typical wave height $a_0$, and the following nondimensionalization of
the independent and dependent variables, cf. \cite{P}, \cite{BCS1}, \cite{C3},
\cite{N}, 
$$
x=\frac{\tilde{x}}{\lambda_0},\,\,\,y=\frac{\tilde{y}}{\lambda_0},\,\,\,
z=\frac{\tilde{z}}{h_0},\,\,\, t=\frac{c_0}{\lambda_0}\tilde{t},$$
and
$$u=\frac{h_0}{a_0c_0}\tilde{u},\,\,\,v=\frac{h_0}{a_0c_0}\tilde{v},
\,\,\, w=\frac{\lambda_0}{a_0c_0}\tilde{w}, \,\,\,
\eta=\frac{\tilde{\eta}}{a_0},\,\,\, h=\frac{\tilde{h}}{h_0},
\,\,\,D=\frac{\tilde{D}}{h_0},\,\,\,\zeta=\frac{\tilde{\zeta}}{a_0},$$
where $c_0=\sqrt{gh_0}$. Then, the governing equations for the fluid motion in 
nondimensional and scaled variables take the following form:
\begin{equation}\label{E2.6}
\varepsilon {\bf u}_t+\varepsilon^2(({\bf u}\cdot\nabla) {\bf
u}+w{\bf u}_z)+\frac{1}{\rho c_0^2}\nabla \tilde{P}=0,
\end{equation}
\begin{equation}\label{E2.7}
\varepsilon\sigma^2 w_t+\varepsilon^2\sigma^2({\bf u}\cdot\nabla
w)+ww_z)+\frac{1}{\rho c_0^2}\tilde{P}_z=-1,
\end{equation}
for $-h<z<\varepsilon\eta$. Here ${\bf u}=(u,v)^T$ and $\nablae=(\partial_x,\partial_y)^T$, 
and the parameters
$\varepsilon=\alpha_0/h_0$ and $\sigma=h_0/\lambda_0$ are assumed
to be small. The conservation of mass is formulated as
\begin{equation}\label{E2.8}
\nabla\cdot {\bf
u}+w_z=0\,\,\,\,\mbox{for}\,\,\,-h<z<\varepsilon\eta,
\end{equation}
while the irrotationality condition is expressed by the equations
\begin{eqnarray}
u_y-v_x=0,& & \label{E2.9} \\
{\bf u}_z-\sigma^2 \nabla w=0, & & \label{E2.10}
\end{eqnarray}
for $-h<z<\varepsilon\eta$; the boundary conditions take the
form
\begin{equation}\label{E2.11}
\eta_t+\varepsilon ({\bf u}\cdot\nabla
\eta)-w=0\,\,\,\mbox{on}\,\, z=\varepsilon\eta,
\end{equation}
\begin{equation}\label{E2.12}
\zeta_t+{\bf u}\cdot \nabla h+w=0\,\,\,\mbox{on}\,\, z=-h.
\end{equation}
We note that $h=D+\varepsilon\zeta$ and thus $h_t=O(\varepsilon)$.

Now, following \cite{P} (see also \cite{DD}), one may derive the generalization of Peregrine's equations with time-dependent variable bottom
\begin{equation}\label{E2.13}
\begin{array}{l}
\eta_t+\nabla\cdot [(h+\varepsilon \eta)\bar{{\bf u}}]+\zeta_t=0,\\
\bar{{\bf u}}_t+\nabla\eta+\varepsilon(\bar{{\bf
u}}\cdot\nabla)\bar{{\bf u}}
-\sigma^2\frac{h}{2}\nabla(\nabla\cdot (h\bar{{\bf
u}}_t))+\sigma^2\frac{h^2}{6}\nabla(\nabla\cdot \bar{{\bf
u}}_t)-\sigma^2\frac{h}{2}\nabla\zeta_{tt}=O(\varepsilon
\sigma^2,\sigma^4).
\end{array}
\end{equation}

Following the methodology of \cite{N} (see also \cite{DD} and \cite{BCS1}), consider the horizontal
velocity of the fluid ${\bf u}^{\theta}$ at some height
$z=-h+\theta(\varepsilon\eta+h)$, with $0\le\theta\le 1$. Then, one may derive from (\ref{E2.6})--(\ref{E2.12}), using appropriate expansions, the generalization of (\ref{E1.6}) given by
\begin{equation}\label{E2.14}
\begin{array}{l}
\eta_t+\nabla\cdot(h{\bf u}^{\theta})+\varepsilon\nabla\cdot(\eta{\bf
u}^{\theta})+\sigma^2\nabla\cdot\left[ \tilde{a}h^2\nabla(\nabla\cdot(h{\bf
u}^{\theta}))+\tilde{b} h^3\nabla(\nabla\cdot {\bf
u}^{\theta})\right]\\\quad +\sigma^2\tilde{a}\nabla\cdot(h^2\nabla\zeta_t)+
\zeta_t=O(\varepsilon\sigma^2,\sigma^4),\\
{\bf u}^{\theta}_t+\nabla\eta+\varepsilon ({\bf u}^{\theta}\cdot \nabla){\bf u}^{\theta}
+\sigma^2\left[\tilde{c}h\nabla(\nabla\cdot(h{\bf u}^{\theta}_t))
+\tilde{d} h^2\nabla(\nabla\cdot {\bf u}^{\theta}_t)
\right]+\sigma^2\tilde{c}h\nabla\zeta_{tt}=O(\varepsilon\sigma^2,\sigma^4),
\end{array}
\end{equation}
where $\tilde{a}=\theta-\frac{1}{2}$, $
\tilde{b}=\frac{1}{2}\left[(\theta-1)^2-\frac{1}{3}\right]$,
$\tilde{c}=\theta-1$ and $\tilde{d}=\frac{1}{2}(\theta-1)^2$.
(We note that the system (\ref{E2.14}) is
Nwogu's system with $z_\alpha=(\theta-1)h$ in the notation of \cite{N} and ${\bf u}={\bf u}^{\theta}+O(\sigma^2)$.)

We observe that due to the irrotationality condition (\ref{E2.9}),
$({\bf u}^{\theta}\cdot \nabla){\bf u}^{\theta}=\frac{1}{2}\nabla |{\bf u}^{\theta}|^2+O(\sigma^2)$, and
$\nabla(\nabla\cdot {\bf u}^{\theta})=\Delta {\bf u}^{\theta}+O(\sigma^2)$. Moreover, there holds
that $h\Delta {\bf u}^{\theta}=\nabla(\nabla\cdot (h{\bf u}^{\theta}))-\nabla(\nabla
h\cdot{\bf u}^{\theta})-\nabla h\nabla\cdot {\bf u}^{\theta}+O(\sigma^2)$, and
$\nabla(\nabla\cdot(h{\bf u}^{\theta}_t))=\nabla(\nabla h\cdot {\bf
u}^{\theta}_t)+\nabla h\nabla\cdot {\bf u}^{\theta}_t+h\Delta {\bf u}^{\theta}_t+O(\sigma^2)$. So,
the system (\ref{E2.14}), after dropping the superscript $\theta$, may be written in the form
\begin{equation}\label{E2.15}
\begin{array}{l}
\eta_t+\nabla\cdot(h{\bf u})+\varepsilon\nabla\cdot(\eta{\bf
u})+\sigma^2\nabla\cdot\left\{(\tilde{a}+\tilde{b})h^2\nabla(\nabla\cdot(h{\bf
u}))-\tilde{b} h^2[\nabla(\nabla h\cdot {\bf u})+\nabla h\nabla
\cdot{\bf
u}]\right\}\\\quad+\sigma^2\tilde{a}\nabla\cdot(h^2\nabla\zeta_t)+
\zeta_t=O(\varepsilon\sigma^2,\sigma^4),\\
{\bf u}_t+\nabla\eta+\varepsilon \frac{1}{2}\nabla |{\bf u}|^2
+\sigma^2\left\{\tilde{c}h[\nabla(\nabla h\cdot{\bf u}_t)+\nabla
h\nabla\cdot{\bf u}_t] +(\tilde{c}+\tilde{d}) h^2 \Delta{\bf u}_t
\right\}\\ \quad+\sigma^2\tilde{c}h\nabla\zeta_{tt}=O(\varepsilon\sigma^2,\sigma^4).
\end{array}
\end{equation}

From these equations we observe that
\begin{equation}\label{E2.16}
\eta_t=-\nabla\cdot (h{\bf u})-\zeta_t+O(\varepsilon,\sigma^2),\,\,\,\,\mbox{and}\,\,\,\,\,
{\bf u}_t=-\nabla \eta+O(\varepsilon,\sigma^2).
\end{equation}
Let $\mu,\nu\in\Rset$. Splitting the dispersive terms using (\ref{E2.16}) (in the spirit of \cite{BCS1}) as follows
\begin{equation}\label{E2.17}
\begin{array}{l}
\nabla(\nabla\cdot(h{\bf u}))=\mu\nabla(\nabla\cdot(h{\bf u})) -(1-\mu)\nabla(\eta_t+\zeta_t)+O(\varepsilon,\sigma^2),\\
\Delta{\bf u}_t=\nu \Delta{\bf u}_t -(1-\nu) \Delta\nabla\eta+O(\varepsilon,\sigma^2),
\end{array}
\end{equation}
and ignoring the high order terms, we may write the system (\ref{E2.15}) in the form
\begin{equation}\label{E2.18}
\begin{array}{l}
\eta_t+\nabla\cdot((h+\varepsilon \eta){\bf u})+\sigma^2
\nabla\cdot\left\{Ah^2[\nabla(\nabla h\cdot{\bf u})+\nabla
h\nabla\cdot {\bf u}]+ ah^2\nabla(\nabla\cdot(h{\bf
u}))-bh^2\nabla
\eta_t\right\}\\
\quad+\sigma^2\tilde{A}\nabla\cdot(h^2\nabla\zeta_t)+
\zeta_t=0,\\
{\bf u}_t+\nabla\eta+\varepsilon \frac{1}{2}\nabla |{\bf u}|^2
+\sigma^2\left\{Bh[\nabla(\nabla h\cdot\nabla\eta)+\nabla
h\Delta\eta] +ch^2\nabla(\Delta\eta)-dh^2\Delta {\bf
u}_t\right\}-\sigma^2Bh\nabla\zeta_{tt}=0,
\end{array}
\end{equation}
and in dimensional form
\begin{equation}\label{E2.19}
\begin{array}{l}
\eta_t+\nabla\cdot((h+ \eta){\bf u})+
\nabla\cdot\left\{Ah^2[\nabla(\nabla h\cdot{\bf u})+\nabla
h\nabla\cdot {\bf u}]+
ah^2\nabla(\nabla\cdot(h{\bf u}))-bh^2\nabla \eta_t\right\}\\
\quad+\tilde{A}\nabla\cdot(h^2\nabla\zeta_t)+
\zeta_t=0,\\
{\bf u}_t+g\nabla\eta+ \frac{1}{2}\nabla |{\bf u}|^2
+\left\{Bgh[\nabla(\nabla h\cdot\nabla\eta)+\nabla h\Delta\eta]
+cgh^2\nabla(\Delta\eta)-dh^2\Delta {\bf
u}_t\right\}-Bh\nabla\zeta_{tt}=0,
\end{array}
\end{equation}
where
$$A=\frac{1}{2}[\frac{1}{3}-(\theta-1)^2],\,\,\,\, B=1-\theta,\,\,\,\,
\tilde{A}=\mu\tilde{a}-(1-\mu)\tilde{b},$$
$$a=\frac{1}{2}\left(\theta^2-\frac{1}{3}\right)\mu,\,\,\,
b=\frac{1}{2}\left(\theta^2-\frac{1}{3}\right)(1-\mu),$$
$$c=\frac{1}{2}\left(1-\theta^2\right)\nu,\,\,\,
d=\frac{1}{2}\left(1-\theta^2\right)(1-\nu).$$

We note that the parameters $a,b,c,d$ are those of the class of Boussinesq systems derived in \cite{BCS1}; if we take the depth $h$ as constant, the systems (\ref{E2.18}) reduce to the analogous systems of
\cite{BCS1}. 

\subsection{Generalization of (\ref{E1.4})}

Following the same procedure as in \cite{BCL} one may generalize (\ref{E1.4}) to a class of Boussinesq systems similar to (\ref{E2.19}). Specifically, considering the shallow water wave equations
\begin{equation}\label{E3.1}
\begin{array}{l}
\eta_t+\nabla\cdot((h+ \varepsilon\eta){\bf u})+\zeta_t=0,\\
{\bf u}_t+g\nabla\eta+ \frac{\varepsilon}{2}\nabla |{\bf u}|^2 =0.
\end{array}
\end{equation}
and using the nonlinear change of variables
$h \tilde{\bf u}={\bf u}(h+\frac{\varepsilon}{2}\eta)$, the fact that 
$\tilde{\bf u}={\bf u}+O(\varepsilon)$, and that $h_t=\varepsilon \zeta_t$,
one may derive in dimensional variables the system
\begin{equation}\label{E3.9}
\begin{array}{l}
\eta_t+\nabla\cdot(h{\bf u})+\frac{1}{2}\nabla\cdot({\bf u}\eta)+
\nabla\cdot\left\{Ah^2[\nabla(\nabla h\cdot{\bf u})+\nabla
h\nabla\cdot {\bf u}]+ ah^2\nabla(\nabla\cdot(h{\bf
u}))-bh^2\nabla
\eta_t\right\}\\
\quad+\tilde{A}\nabla\cdot(h^2\nabla\zeta_t)+
\zeta_t=0,\\
{\bf u}_t+g\nabla\eta+\frac{1}{2h}g\eta\nabla\eta+ \frac{1}{2}
\nabla |{\bf u}|^2+\frac{1}{2h}{\bf u}\nabla\cdot (h{\bf
u})+\frac{1}{2h}u\zeta_t\\
\quad+\left\{Bg[\nabla(\nabla h\cdot\nabla\eta)+\nabla
h\Delta\eta] +cgh\nabla(\Delta\eta)-dh\Delta {\bf
u}_t\right\}-\sigma^2B\nabla\zeta_{tt}=0,
\end{array}
\end{equation}
where $A,\tilde{A},B,a,b,c,d$ as before. If we neglect the dispersive terms of (\ref{E3.9}), then the resulting system conserves the 
function $E(t)=\int_{\Rset^2}
g\eta^2+h|{\bf u}|^2$, in the sense that $E(t)=E(0)$. Moreover, if we choose constant depth $h$, we recover the fully symmetric
Boussinesq systems derived in \cite{BCL}. As Peregrine pointed out in \cite{P}, $h$ and its derivatives must be of $O(1)$ to ensure the validity of all the above models. For other symmetric Boussinesq systems over variable bottom we refer to \cite{Chazel}.

{\bf Remark:} For specific examples of systems produced by specific choices of the parameters $\mu$, $\nu$, $\theta$ we refer to \cite{BCS1}, \cite{BCS2}, \cite{BC}, \cite{DMreview}. We mention that other triplets of parameters may be chosen so that the dispersion relation of the Boussinesq system approximates better the dispersion relation of the full water wave problem. Such a choice is for example, $\mu=0$, $\nu=(25-\sqrt{1605})/49$ and $\theta^2=(80-\sqrt{1605})/105$,
i.e. $a=0$, $b\cong 0.0235121$, $c\cong -0.0952381$, $d\cong 0.4050592$, which gives a (2,4)-Pad\'{e} approximant of the full water wave problem. Another choice is $\mu=-0.3672365$, $\nu=-0.3301459$ and
$\theta^2=0.3906251$ (i.e., $a= -0.0105198$, $b=0.0391657$, $c=-0.1005913$,
$d=0.4052788$), which gives a (4,4)-Pad\'{e} approximant of the full water wave problem. 

\section{The simplified system and the numerical method}

In computations we have used a {\em simplified} Boussinesq system of BBM type.
Consider the system (\ref{E2.19}) with $\theta^2=2/3$ and $\mu=\nu=0$, i.e., $A=\tilde{A}=\sqrt{\frac{2}{3}}-\frac{2}{3}$, $B=1-\sqrt{\frac{2}{3}}$, $a=c=0$, $b=d=1/6$. Then assuming that the bottom 
$h$ is such that the derivatives of order greater one are of $O(\varepsilon)$ and omitting terms of $O(\varepsilon^2,\varepsilon\sigma^2)$, we consider the following
initial-boundary-value problem for the system (\ref{E2.19}) for $x$ in a plane bounded domain $\Omega$
and $t\geq 0$: 
\begin{equation}\label{E4.1}
\begin{array}{l}
\eta_t+\nabla\cdot((D+\zeta+ \eta){\bf u})+2AD\nabla D\cdot
\nabla(\nabla\cdot(D{\bf u}))
-b\nabla\cdot(D^2\nabla \eta_t)+\tilde{A}\nabla\cdot(D^2\nabla\zeta_t)+\zeta_t=0,\\
{\bf u}_t+g\nabla\eta+ \frac{1}{2}\nabla |{\bf u}|^2 +BgD[(\nabla
D\cdot\nabla)\nabla\eta)+\nabla D\Delta\eta] -dD^2\Delta {\bf
u}_t-BD\nabla\zeta_{tt}=0,
\end{array}
\end{equation}
$$\eta(x,y,0)=\eta_0(x,y),\,\,
u(x,y,0)=u_0(x,y),\,\,v(x,y,0)=v_0(x,y),\,\,(x,y)\in\overline{\Omega},$$
with Dirichlet boundary conditions $\eta=u=v=0$ on $\partial\Omega$, for $t\geq 0$.

We solved the above initial-boundary-value problem using the standard Galerkin-finite element
method with continuous P1 elements on a triangulation of $\Omega$. For the time stepping we used an explicit
Runge-Kutta method of order 2 with a uniform timestep. This
numerical scheme was analyzed and used in \cite{DMS1} and
\cite{DMS2}, in the case of the BBM-BBM Boussinesq system with constant
depth. The main difference in the case of a general variable
bottom is that the matrices that the semidiscretization of the
${\bf u}$-equations yields are not symmetric, and thus for the
numerical solution of the linear systems for $u$ and $v$ we use
the Generalized Minimal Residual Method (GMRES) with appropriate
ILUT preconditioner, cf. \cite{Saad}. For the bottom $h$ we use the
interpolant in the finite element space $V_{\rm h}$, while we approximate the
initial data using an appropriate elliptic projection into
$V_{\rm h}$. For example, the initial condition for the $u$
component of the solution is the function $u_{0,{\rm h}}\in V_{\rm h}$
which satisfies
\begin{equation*}
{\cal A}(u_{0,{\rm h}},\chi)={\cal A}(u_0,\chi),\,\,\,\mbox{for all}\,\,\,\chi\in V_{\rm h},
\end{equation*}
where ${\cal A}$ is the bilinear form defined by the relation
\begin{equation*}
{\cal A}(u,v)=(u,v)+d(D^2\nabla u,\nabla v)+d(\nabla D^2 \nabla
u,v),\,\,\,\mbox{for all}\,\,\,u,v\in H^1_0,
\end{equation*}
where $H^1_0$ is the usual Sobolev space $W^{1,2}_0(\Omega)$.

From the previous comments it is obvious that we will have to impose restrictions on the smoothness of the bottom topography to ensure the well-posedness of (\ref{E4.1}) and the stability of the numerical method. For the triangulation of the domain $\Omega$ we usually use the ``Triangle'' software, \cite{Triangle}.

\section{Tsunami generation phase}\label{tg}

Many tsunami waves are caused by a bottom dislocation near a
rupture due to an earthquake. In that case the bottom deformation
may be approximated by Okada's formulas, \cite{Ok1}, \cite{Ok2}. We
briefly describe the vertical component of Okada's formulas in the case of the so called dip-slip dislocation, which is the most important component in the case of tsunami generation.

In this case, consider a rectangular fault of width $W$ and length $L$ positioned near the $z\le 0$ axis and at a depth $d$ below the free surface
$z=0$. The vector $D$ represents the slip on the fault. The dip
angle $\delta$ and the angle $\phi$ between the the fault plane
and the slip vector, describe a general dislocation of the
fault, where the vertical component of the displacement vector, (which we will denote here by ${\cal O}(x,y)$), is given by the following formulas in Chinnery's notation, 
$\left.f(\xi,\eta)\right\|:=f(y+\frac{L}{2},p)-f(y+\frac{L}{2},p-W)-
f(y-\frac{L}{2},p)+f(y-\frac{L}{2},p-W)$, cf. \cite{Ok1}, \cite{DD1}.

\begin{equation}\label{E6.1}
{\cal O}(x,y)=-\frac{U}{2\pi}\left.\left(\frac{\tilde{d}q}{R(R+\xi)}+
\sin\delta\arctan\frac{\xi\eta}{qR}-I\sin\delta\cos\delta\right)\right\|,
\end{equation}
where $U=|D|\sin\phi$, $p=y\cos\delta+d\sin\delta$,
$q=y\sin\delta-d\cos\delta$,
$\tilde{y}=\eta\cos\delta+q\sin\delta$,
$\tilde{d}=\eta\sin\delta-q\cos\delta$,
$R^2=\xi^2+\eta^2+q^2=\xi^2+\tilde{y}^2+\tilde{d}^2$,
$X^2=\xi^2+q^2$. When $\cos\delta\not=0$, $I$ is given by the
formula
\begin{equation*}
I=\frac{\mu}{\mu+\lambda}\frac{2}{\cos\delta}\arctan
\frac{\eta(X+q\cos\delta)+X(R+X)\sin\delta}{\xi(R+X)\cos\delta},
\end{equation*}
and, when $\cos\delta=0$, by
\begin{equation*}
I=\frac{\mu}{\mu+\lambda}\frac{\xi\sin\delta}{R+\tilde{d}}.
\end{equation*}
Here $\mu$, $\lambda$ are the Lam\'{e} constants given
by the formulas
$\mu=E/2(1+\nu)$, $\lambda=E\nu/(1+\nu)(1-2\nu)$.
The parameter $E$ is the Young's modulus and $\nu$
is the Poisson's ratio; both are considered to be known
constants. In the sequel, it is supposed that Okada's vertical deformation component 
${\cal O}(x,y)$ has been translated to the appropriate location.

In \cite{DD1} and \cite{Ham1}, some mechanisms of the dynamics of tsunami
generation are described. We review here two cases. A common practice in the
literature is to use as initial condition for the free surface
elevation Okada's solution, while the
initial velocity profile is considered to be zero.  
This is referred to as {\em
passive} generation of a tsunami and is described by the
initial conditions
\begin{equation}\label{E6.2}
\eta(x,y,0)={\cal O}(x,y),\,\,\,\,{\bf u}(x,y,0)=0.
\end{equation}

Another way to model the generation of a tsunami is by {\em active} generation. In this case we consider zero initial
conditions for both the surface elevation and velocity field, and
assume that the bottom is changing in time. This case may be described by considering the bottom motion formula
\begin{equation}\label{E6.3}
h(x,y,t)=D(x,y)+{\cal O}(x,y){\cal F}(t),
\end{equation}
where ${\cal F}$ is a function of time $t$. In this paper we
will consider three cases used in \cite{DD1}, \cite{KDD},
\cite{Ham1}.
\begin{equation}\label{E6.4}
\begin{array}{l}
{\cal F}_i(t):={\cal H}(t),\\
{\cal F}_e(t):=1-\exp^{-\kappa t},\,\,\,\kappa>0\\
{\cal F}_c(t):={\cal H}(t-t_0)+\frac{1}{2}[1-\cos(\pi t/t_0)]{\cal
H}(t_0-t).
\end{array}
\end{equation}
In the numerical experiments that follow we use $\kappa=2$ and $t_0=2$sec in the above. 

In each of the above cases the analytical solution of the
linearized Euler equations has been computed in 
\cite{DD1} and is given by the formulas (see also \cite{Ham1}):
\begin{equation}\label{E6.5}
\begin{array}{l}
\eta_i(x,y,t)=\frac{1}{(2\pi)^2}\int_{\Rset^2}
\frac{\hat{\zeta}(k,\ell)e^{{\rm
i}(kx+\ell y)}}{\cosh(mh)}\cos\omega tdkd\ell,\\
\eta_e(x,y,t)=\frac{-\kappa^2}{(2\pi)^2}\int_{\Rset^2}
\frac{\hat{\zeta}(k,\ell)e^{{\rm i}(kx+\ell
y)}}{\cosh(mh)}\left(\frac{e^{-\kappa t}-\cos\omega t
-\frac{\omega}{\kappa}\sin\omega t}{\kappa^2+\omega^2}\right)dkd\ell,\\
\eta_c(x,y,t)=\frac{\gamma^2}{(2\pi)^2}\int_{\Rset^2}
\frac{\hat{\zeta}(k,\ell)e^{{\rm i}(kx+\ell y)}}{\cosh(mh)} \left(
\sin \omega t-\cos\gamma t+{\cal H}(t-t_0)[\cos\omega
(t-t_0)+\cos\gamma t]\right) dkd\ell,
\end{array}
\end{equation}
where $\gamma^2=\frac{\pi}{t_0}$, $m=\sqrt{k^2+\ell^2}$,
$\omega=\sqrt{gm\tanh m h}$. The symbol $\hat{\zeta}$ represents
the Fourier transform of $\zeta(x,y,t)$, i.e.
\begin{equation*}
\hat{\zeta}(k,\ell)=\int_{\Rset^2}\zeta(x,y,t)e^{-{\rm i}(kx+\ell y)}dxdy.
\end{equation*}
In the numerical experiments we evaluated numerically the above formulas by
approximating the usual Fourier transform by the discrete Fourier transform, using the FFT.

\begin{figure}[p]
\begin{center}
\begin{tabular}{cc}
\includegraphics[scale=.266,angle=0]{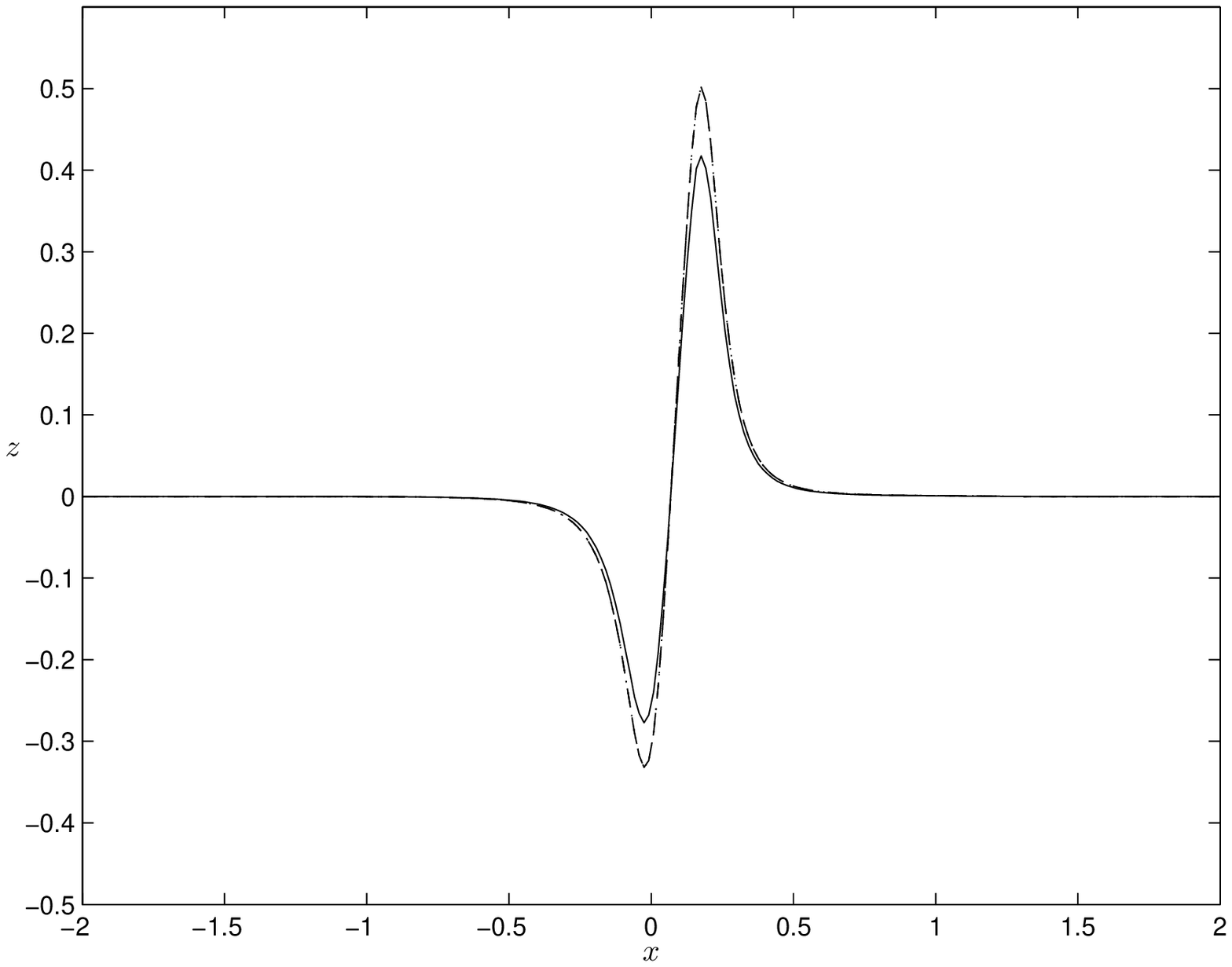} &
\includegraphics[scale=.266,angle=0]{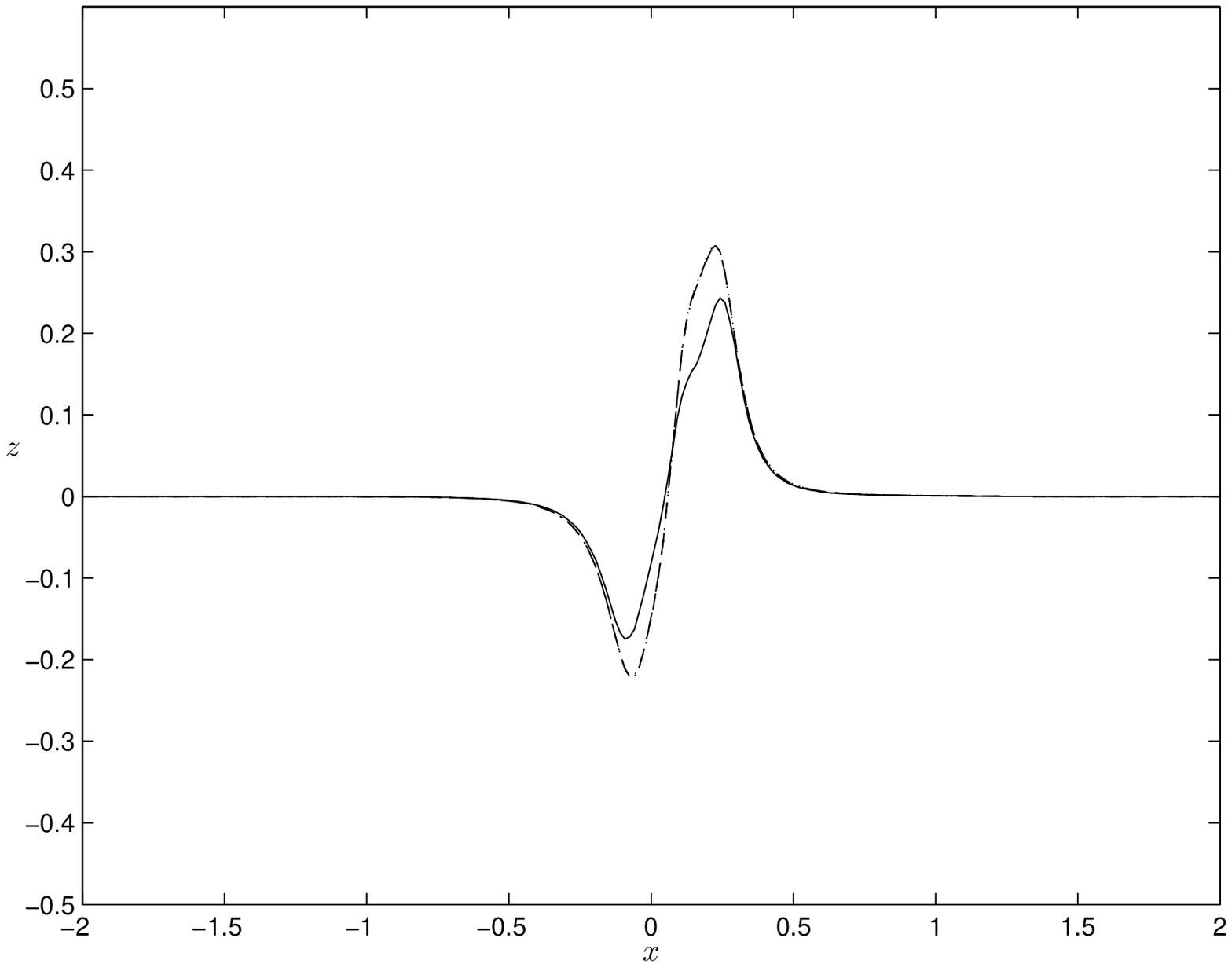}\\
$t=10$sec & $t=100$sec \\
\includegraphics[scale=.266,angle=0]{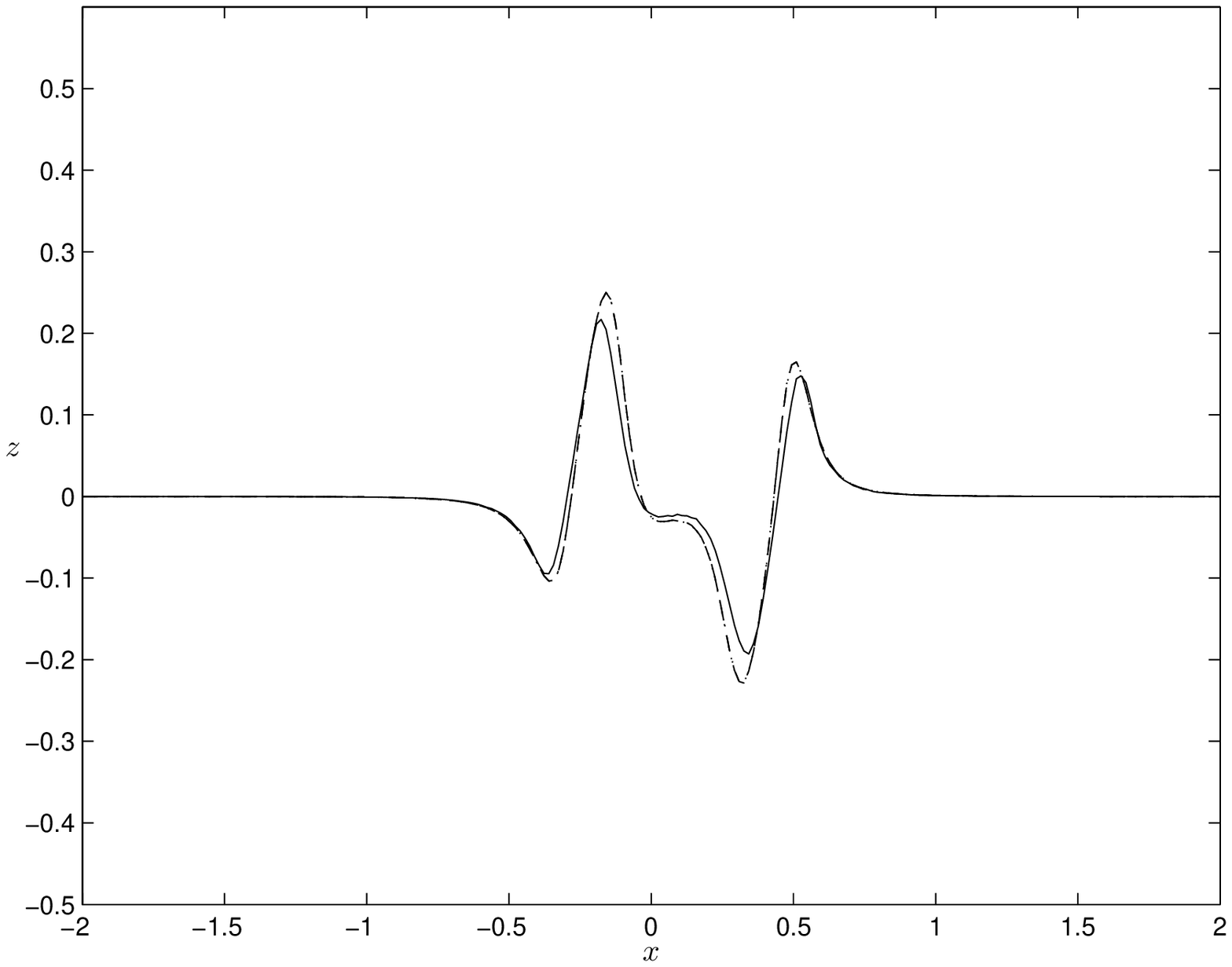}&
\includegraphics[scale=.266,angle=0]{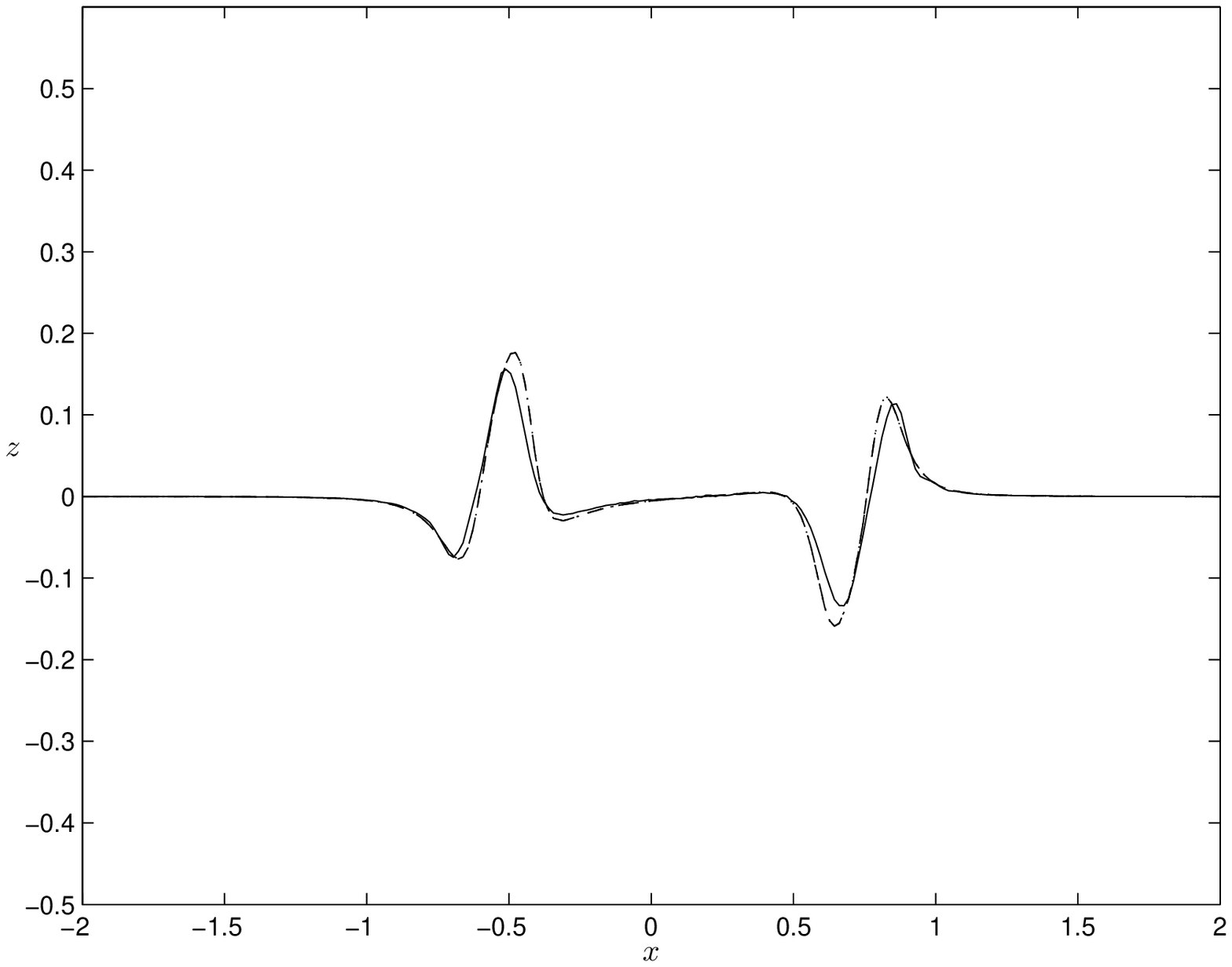}\\
$t=500$sec  & $t=1000sec $
\end{tabular}
\end{center}
\caption{$D_0=500$m, case ${\cal F}_e$, $\kappa=2$. Boussinesq ---,
Euler ${\cal F}_e$ $-\,-$, Euler ${\cal F}_c$ $-\cdot-$, Euler
${\cal F}_i$ $\cdots$}\label{F6.2}
\end{figure}

\begin{figure}[p]
\begin{center}
\begin{tabular}{cc}
\includegraphics[scale=.266,angle=0]{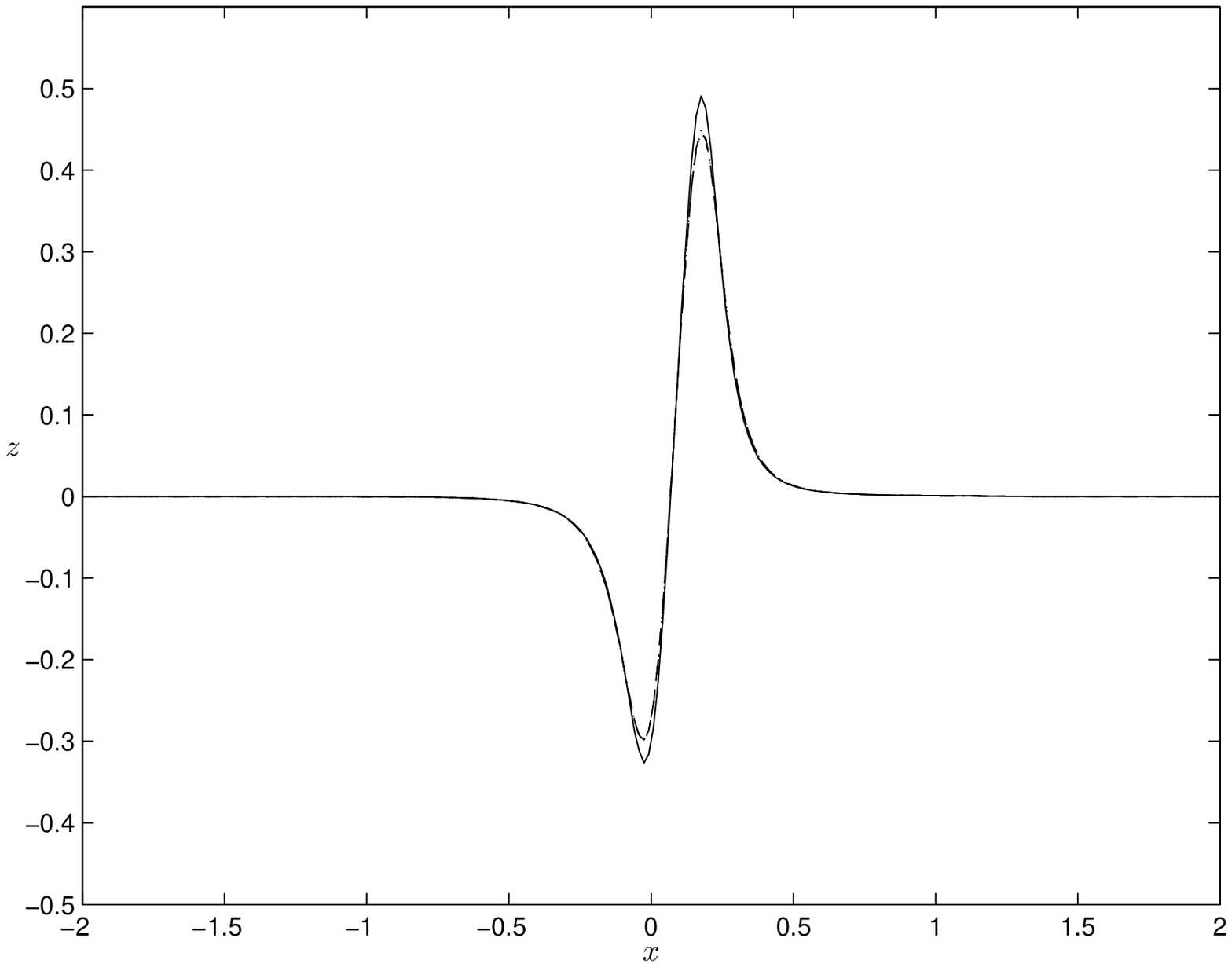} &
\includegraphics[scale=.266,angle=0]{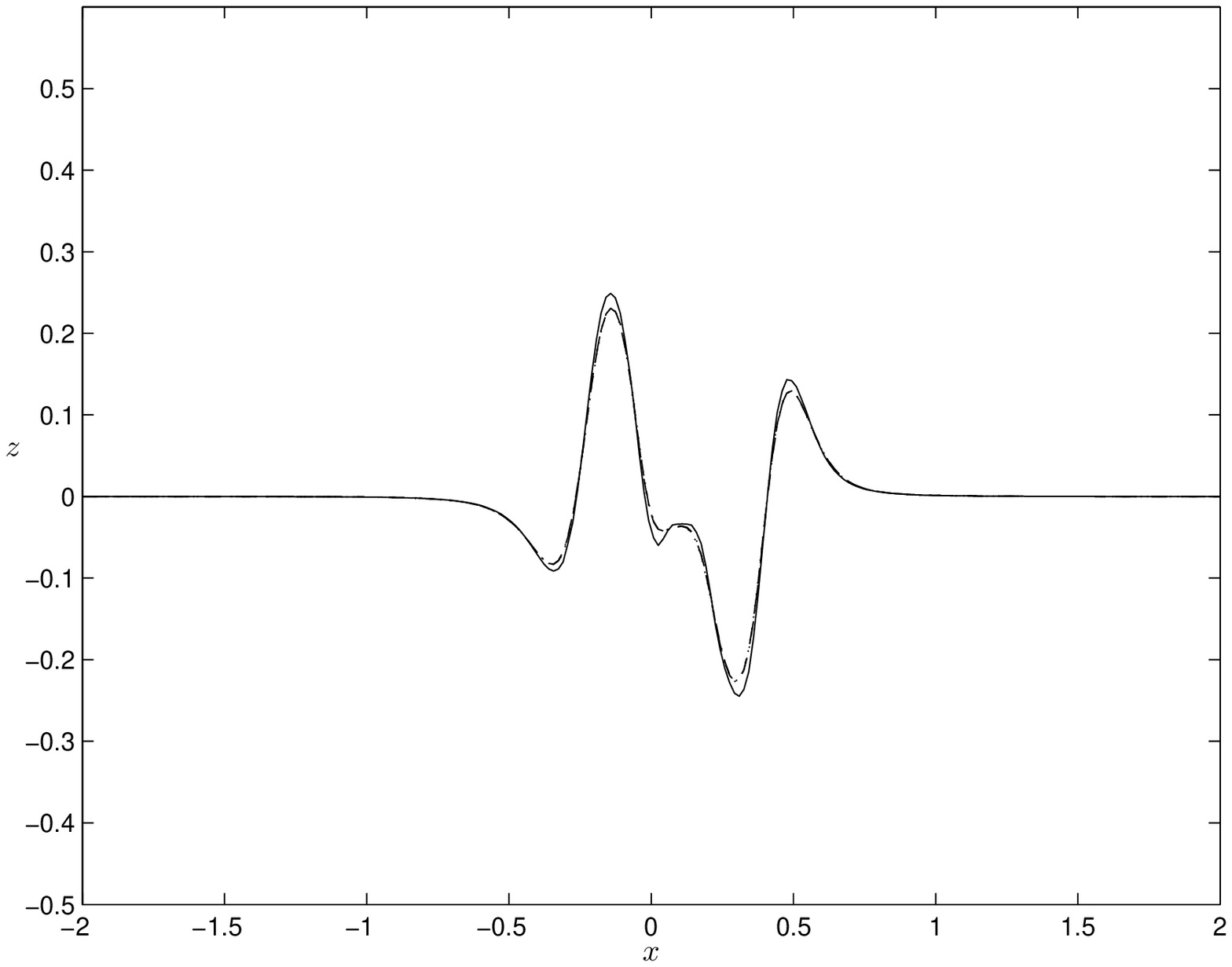}\\
$t=10$sec & $t=200$sec \\
\includegraphics[scale=.266,angle=0]{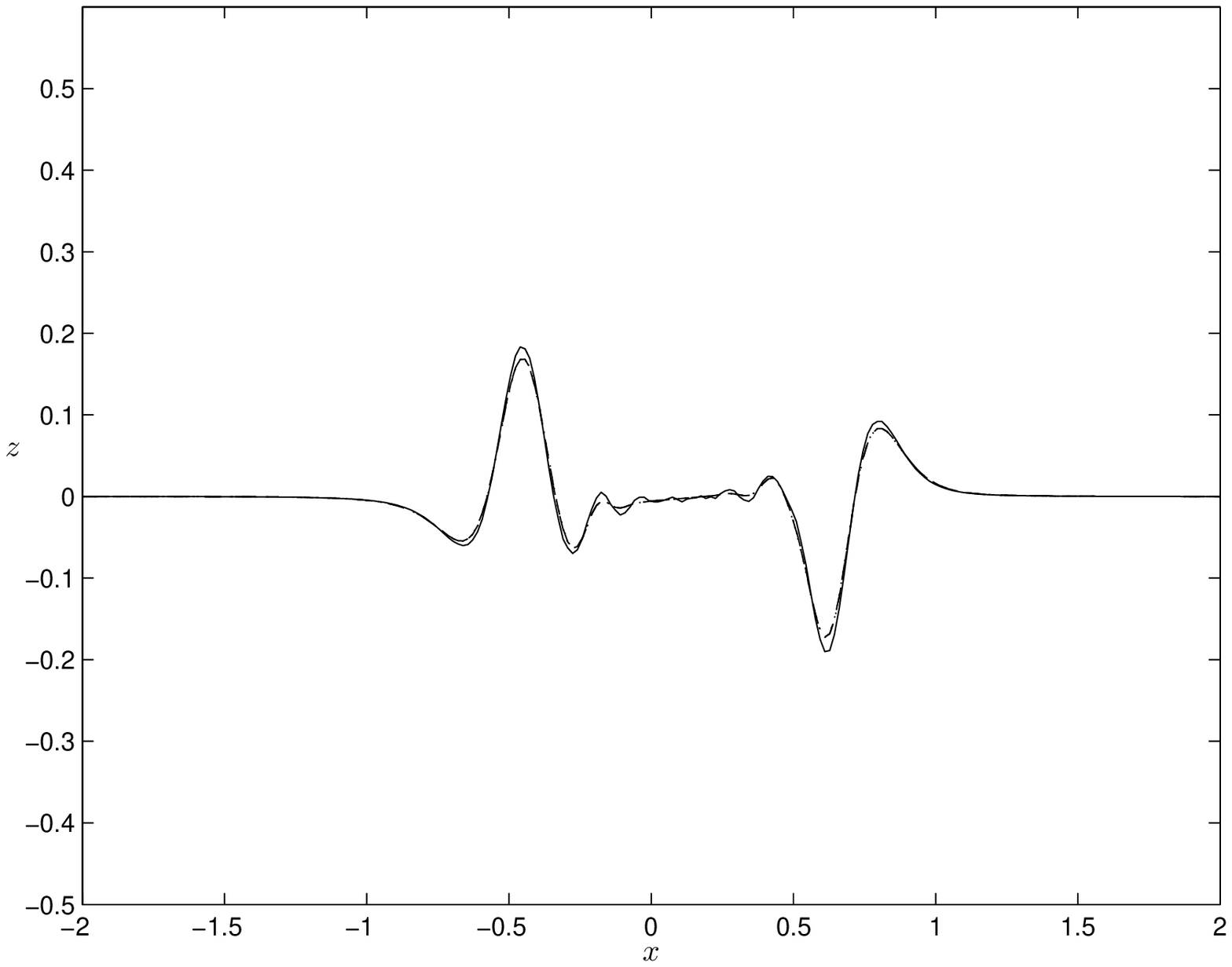}&
\includegraphics[scale=.266,angle=0]{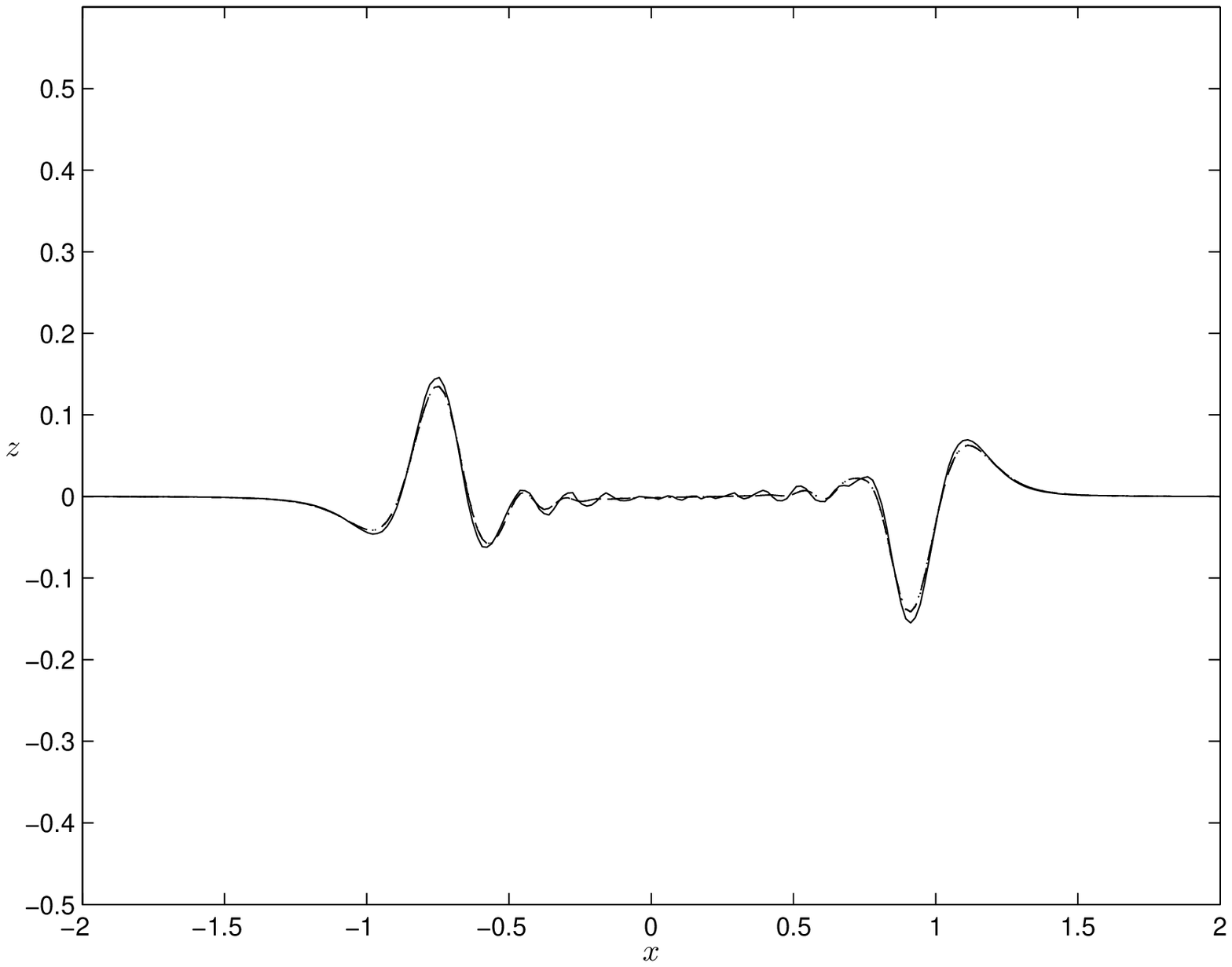}\\
$t=400$sec  & $t=600$sec \\
\end{tabular}
\end{center}
\caption{$D_0=3000$m, case ${\cal F}_c$, $t_0=2$. Boussinesq ---, Euler
${\cal F}_e$ $-\,-$, Euler ${\cal F}_c$ $-\cdot-$, Euler ${\cal
F}_i$ $\cdots$}\label{F6.3}
\end{figure}

\begin{figure}[ht]
\begin{center}
\begin{tabular}{cc}
\includegraphics[scale=.266,angle=0]{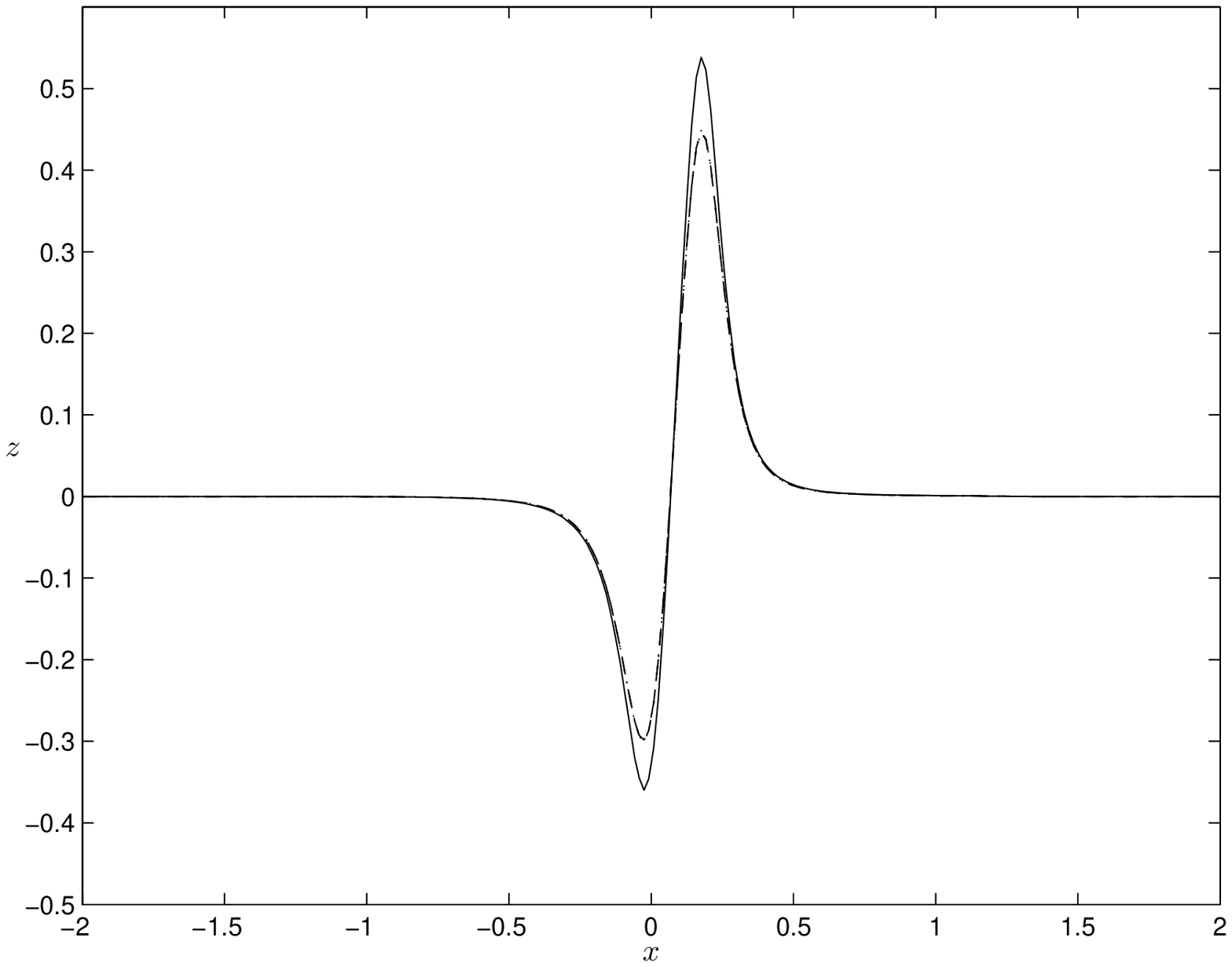} &
\includegraphics[scale=.266,angle=0]{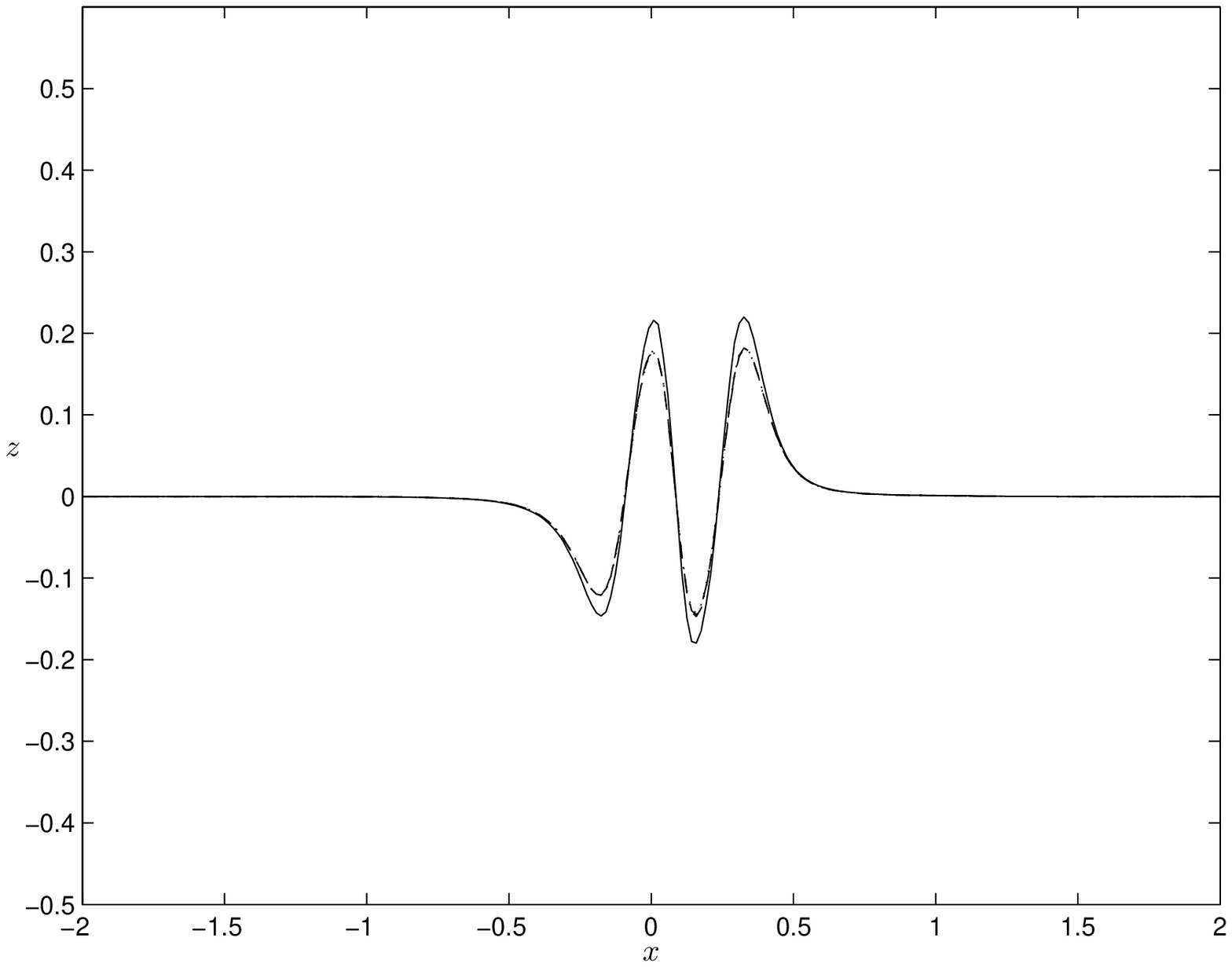}\\
$t=10$sec  & $t=100$sec \\
\includegraphics[scale=.266,angle=0]{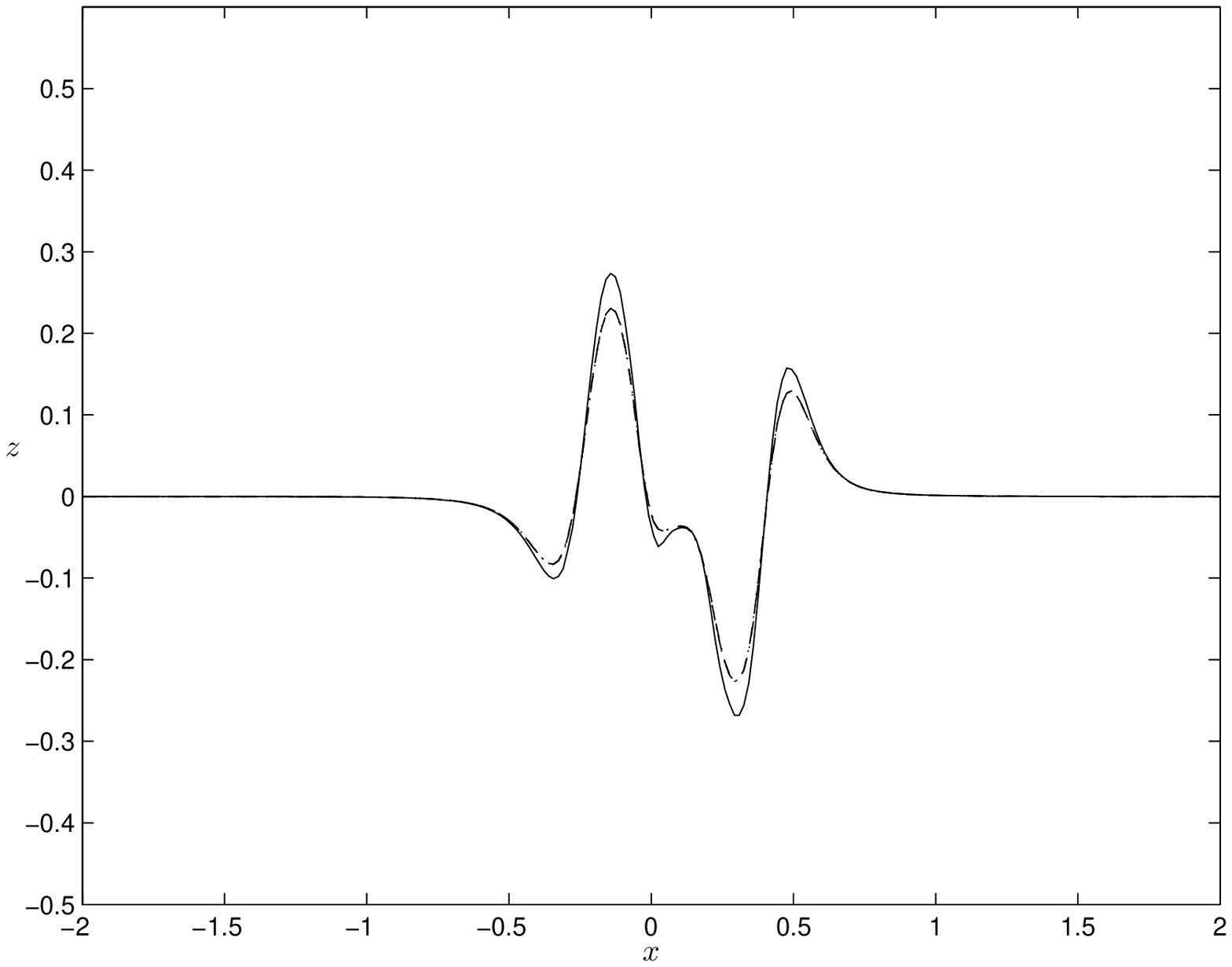}&
\includegraphics[scale=.266,angle=0]{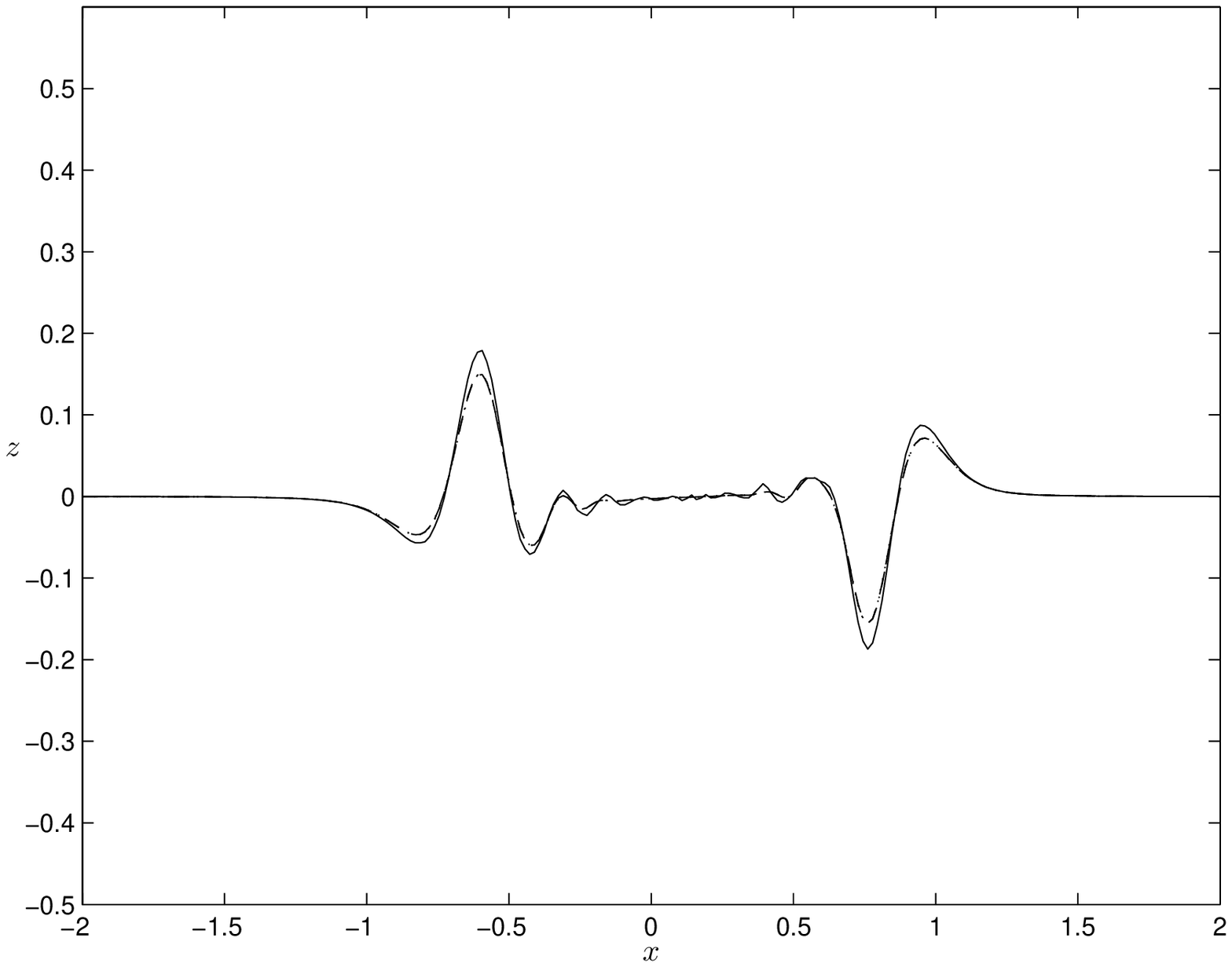}\\
$t=200$sec  & $t=500$sec \\
\end{tabular}
\end{center}
\caption{$D_0=3000$m, case ${\cal F}_i$. Boussinesq ---, Euler ${\cal
F}_e$ $-\,-$, Euler ${\cal F}_c$ $-\cdot-$, Euler ${\cal F}_i$
$\cdots$}\label{F6.4}
\end{figure}

In the sequel, we present a comparison between the numerical solution of the Boussinesq
model (\ref{E4.1}) and the analytical solution of the Euler equations
(\ref{E6.5}) to study the generation of a tsunami. For the deformation of the bottom we used the formulas (\ref{E6.1})--(\ref{E6.4}) in the square $[-2,2]\times [-2,2]$ in geographical coordinates (i.e. longitude-latitude coordinates in degrees).
More precisely, we considered the bottom motion given by $h(x,y,t)=D(x,y)+{\cal O}(x,y){\cal F}(t)$, 
where ${\cal F}$ is one of the functions (\ref{E6.4}), and $D(x,y)=D_0$, with $D_0=500$, $1000$ and $3000$m. 
To compute the vertical displacement function ${\cal O}(x,y)$ from (\ref{E6.1}) we use the set 
of parameters shown in Table \ref{T6.1}.  
Some comparison results appear in figures \ref{F6.2}--\ref{F6.4}. These figures show the surface elevation produced by the two models as a function of spatial variable $x$ when $y=0$ in three cases. The horizontal scales in these figures are in geographical coordinates (degrees). The spherical shape of the earth was taken into account, even if it does not play significant role because of the small spatial scale of the experiments, (near the equator one degree is approximately equal to 111km.).


\begin{table}[ht]
\caption{Parameters used for the first experiment}\label{T6.1}
\begin{tabular}{lc|lc}
\hline Parameter & Value & Parameter & Value \\
\hline
Dip angle $\delta$(deg) & 7.0 & Fault length $L$ (km) & 40 \\
Rake angle $\phi$(deg) & 67 & Fault width $W$ (km) & 20\\
Strike angle (deg) & 90.0 & Fault depth $d$ (km) & 10\\
Slip amount, $|D|$(m) & 2 & Young modulus $E$ (GPa) & 9.5\\
Longitude (deg) & 0 & Poisson ration $\nu$ & 0.27\\
Latitude (deg) & 0 & Acceleration of gravity $g$ (${\rm m/sec^2}$) & 9.81\\
\hline
\end{tabular}
\end{table}

For the finite element code we used $90970$ elements, while for the computation of the FFT we used a rectangular grid of $57600$ squares.
(In these experiments we chose the parameters such that both models
are valid. Recall that the Boussinesq model
is valid when the Stokes number $S$ is $O(1)$. In
practice this means that $0.5\le S\le 35$, cf. \cite{Bona}. For the
choice of the time scales in formulas (\ref{E6.4}), cf. \cite{Ham1}.)

We observe that the solution of the Boussinesq model is close to the analytical solution of the linearized Euler
equations (\ref{E6.5}). The best agreement is apparently achieved in the cases where we used ${\cal F}_c$ and ${\cal F}_i$. For a more thorough study of  tsunami generation we refer to \cite{DD2}.

\section{A case of tsunami propagation: Comparison between MOST
and the Boussinesq code}\label{tp}

In this section we present the results of a simulation of the propagation of the tsunami wave
that affected the island of Java in July 17, 2006, using the Boussinesq
model (\ref{E4.1}) and the MOST code. MOST is an efficient numerical model solving the nonlinear shallow water wave equations in a characteristic form, using a splitting technique, coupled with an explicit second-order in space and first-order in time finite-difference scheme, \cite{Tphd}, \cite{TS}, \cite{TS2}. 

The seismological data that we used for the tsunami generation were
provided by the National Oceanic and Atmospheric Administration - NOAA\footnote{http://www.pmel.noaa.gov}. The fault's length and width were 100km and 50km, respectively, the strike angle was taken equal to $199$deg, while the epicenter was placed at $(107.18, -9.56)$ in geographical coordinates. (Poisson's ratio was $\nu=0.23$ in this case.) For the bathymetric data we used the GEBCO one-minute grid provided by the British Oceanographic Data Centre - BODC\footnote{http://www.bodc.ac.uk/projects/gebco.html}. For the definition of the coastlines which represent part of the boundary of the domain $\Omega$, we used the Global Self-Consistent Hierarchical High-Resolution Shoreline data, GSHHS\footnote{http://www.ngdc.noaa.gov/mgg/shorelines/gshhs.html}. 

\begin{figure}[ht]
\begin{center}
\begin{tabular}{cc}
\includegraphics[scale=.40,angle=0]{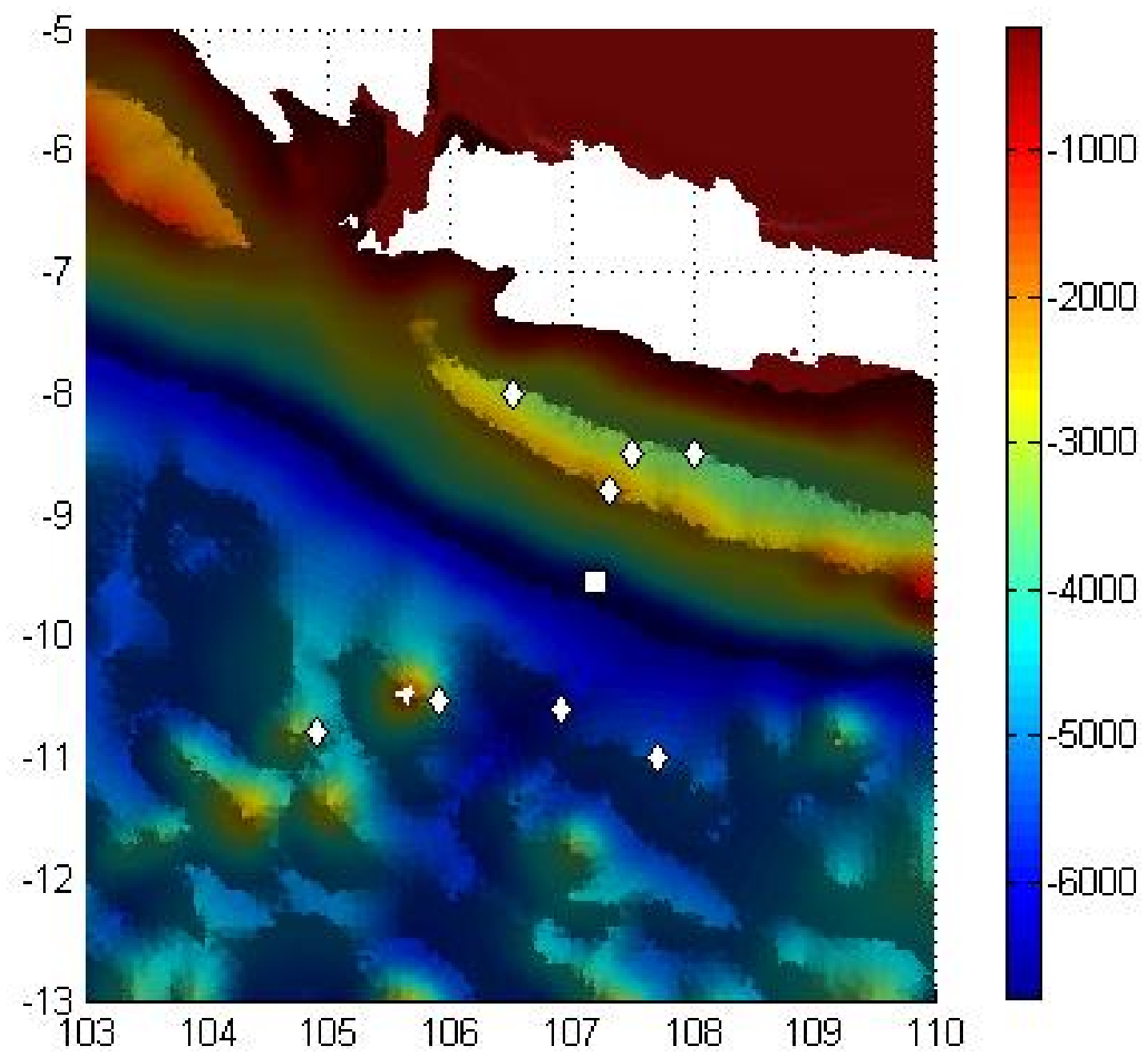}
\end{tabular}
\end{center}
\caption{Bathymetry of the sea around the island of Java used by the Boussinesq model, and position of the wave gauges and the epicenter of the earthquake. $\diamondsuit$: wave gauge, $\square$: the epicenter. (All distances in degrees.)}\label{F8.1}
\end{figure}

In the case of the Boussinesq model, the bathymetric data were interpolated and
smoothed appropriately to ensure the stability of the numerical method and the validity of the Boussinesq system. 
However, because of the large variations of the bottom (see Figure \ref{F8.1}), shorter waves were generated, especially around Christmas Island (southwest of Java) and around the undersea canyon near the earthquake's epicenter. In addition, due to some incompatibility between the shoreline and the bathymetric data we ignored riffs and islands that were not included in the shoreline data by changing the depth close to the shoreline to 100m. In the case of MOST we used the original bathymetric data. (We note that MOST defines the coastlines by an internal procedure when a depth less than 10m is detected.)
In both models we used the passive approach for the
generation of the tsunami, i.e., initial
conditions (\ref{E6.2}). The time step for both models was taken equal to 1sec. For the Boussinesq code we used 85863 elements while for MOST we used a uniform grid of width equal to $0.01$deg. (We ran the same experiment using the Boussinesq code with 80628 and 181480 elements with no significant differences.)

\begin{figure}[ht]
\begin{center}
\begin{tabular}{cc}
\includegraphics[scale=.28,angle=0]{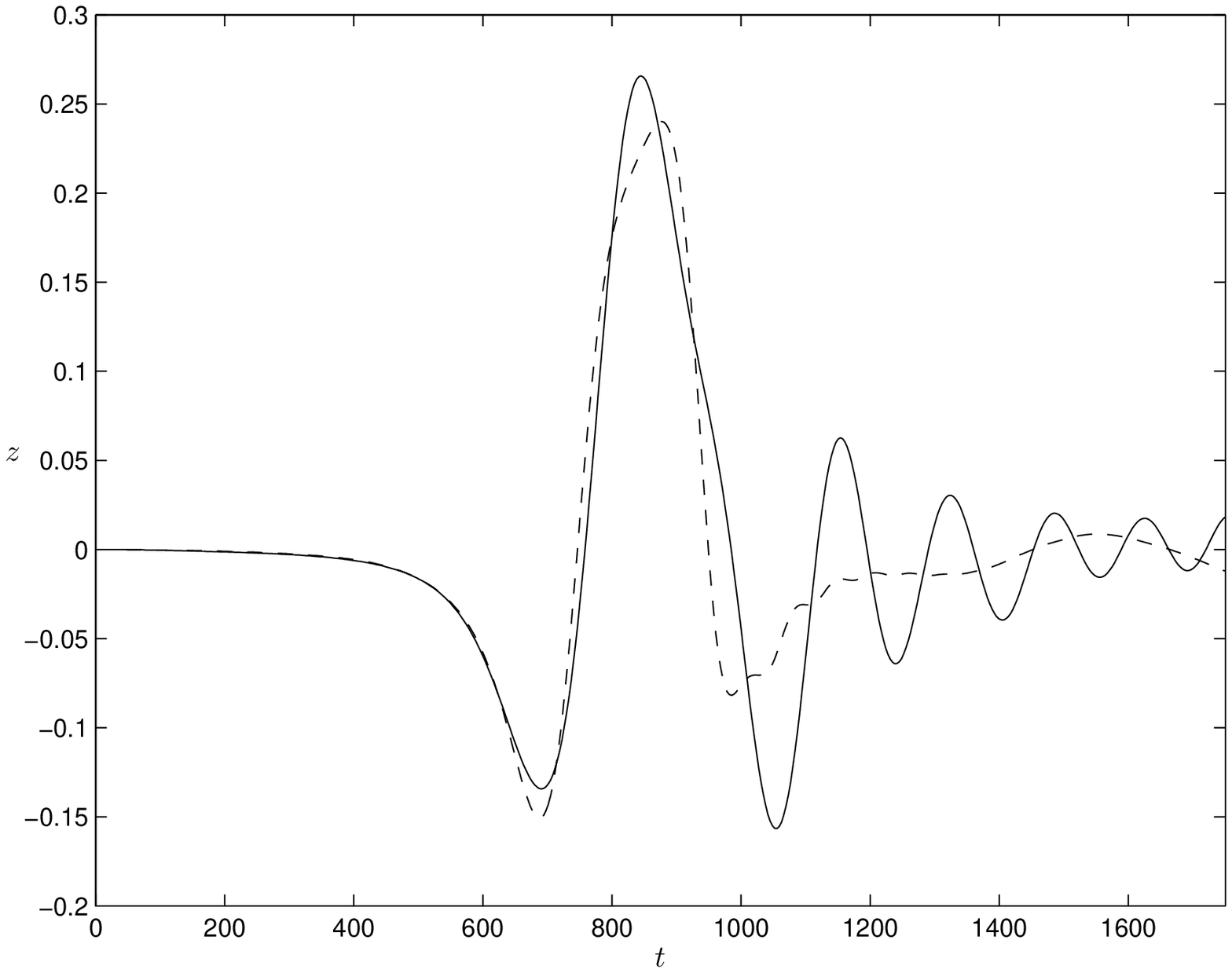}&
\includegraphics[scale=.28,angle=0]{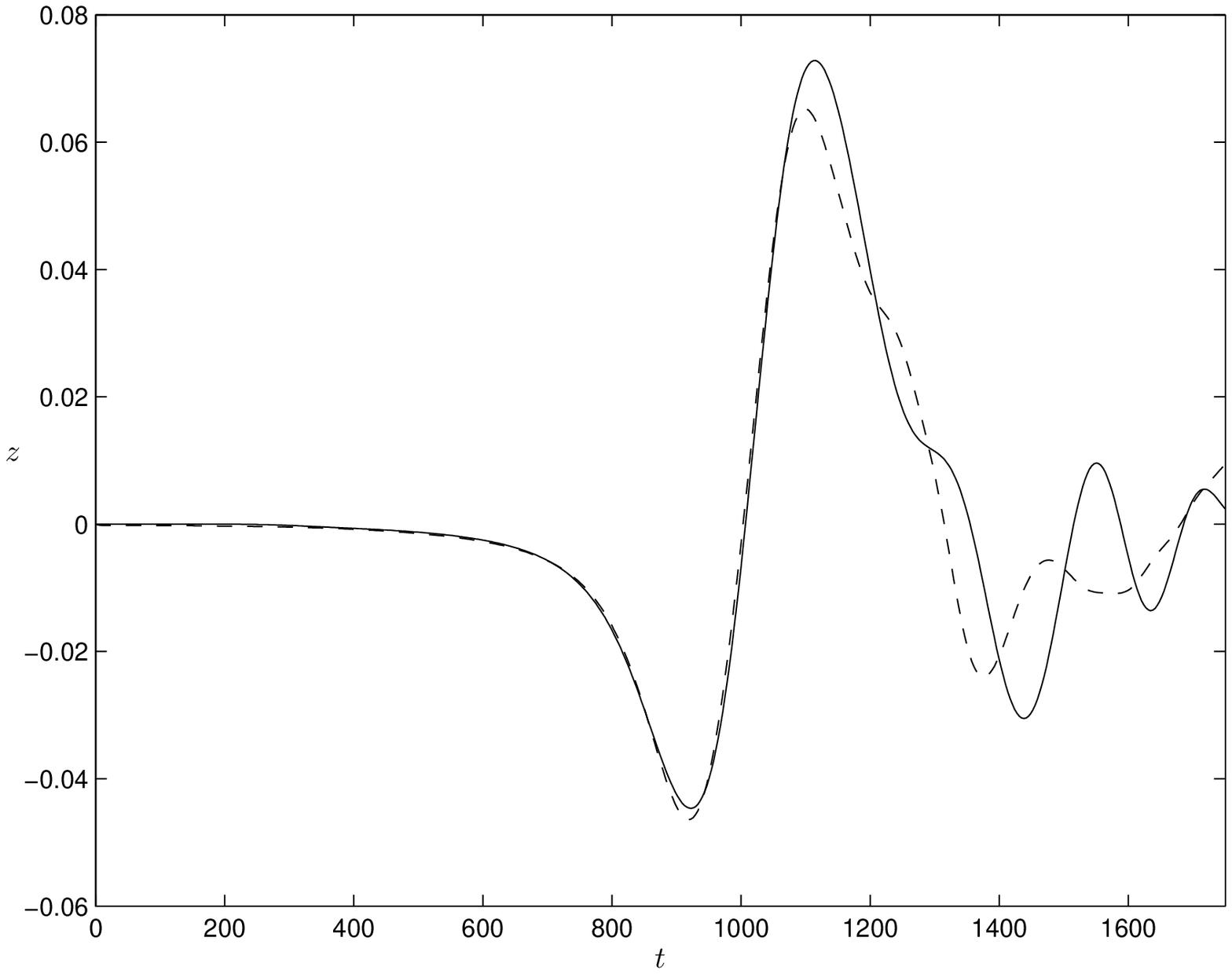}\\
Gauge (i) & Gauge (ii)\\
\includegraphics[scale=.28,angle=0]{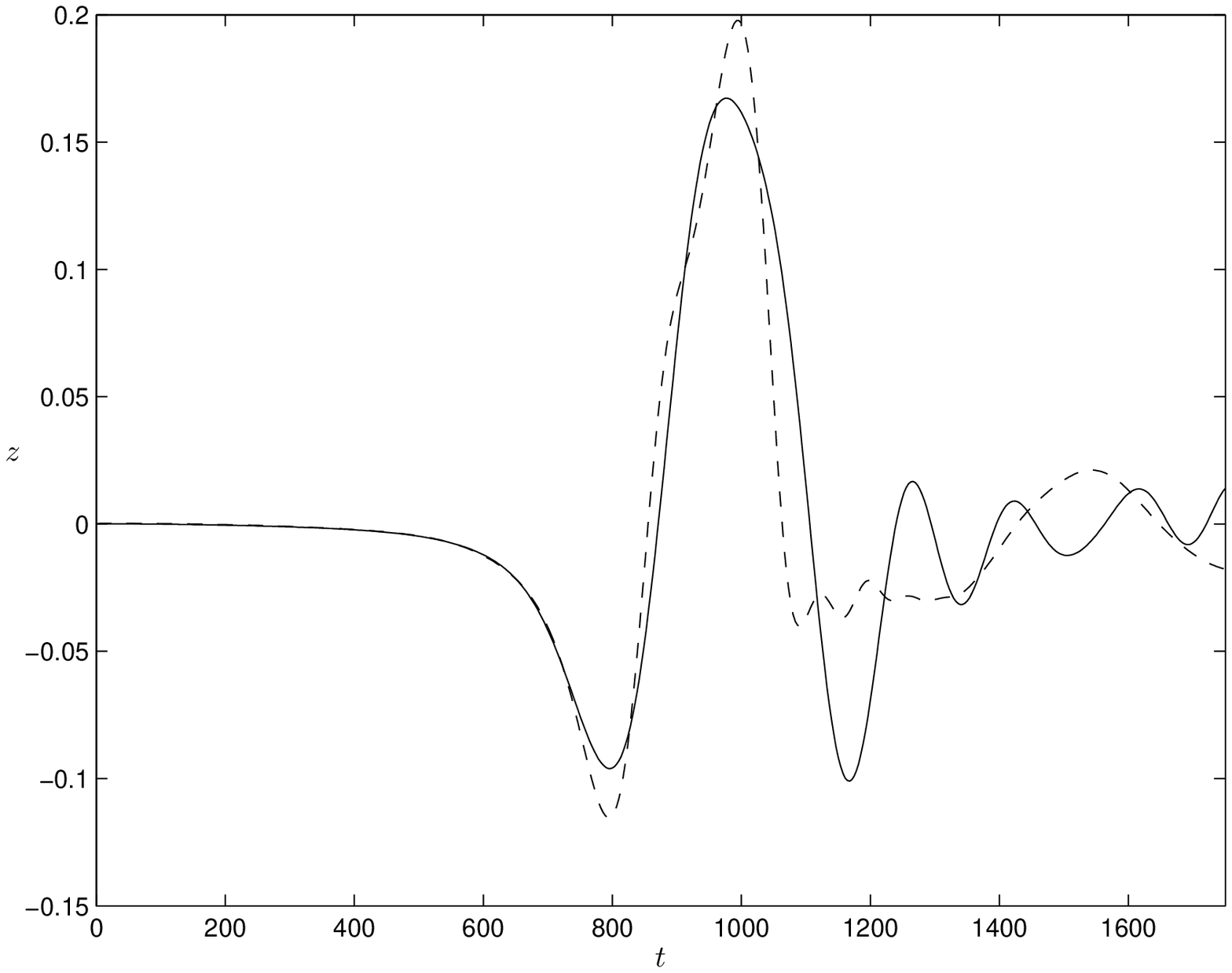}&
\includegraphics[scale=.28,angle=0]{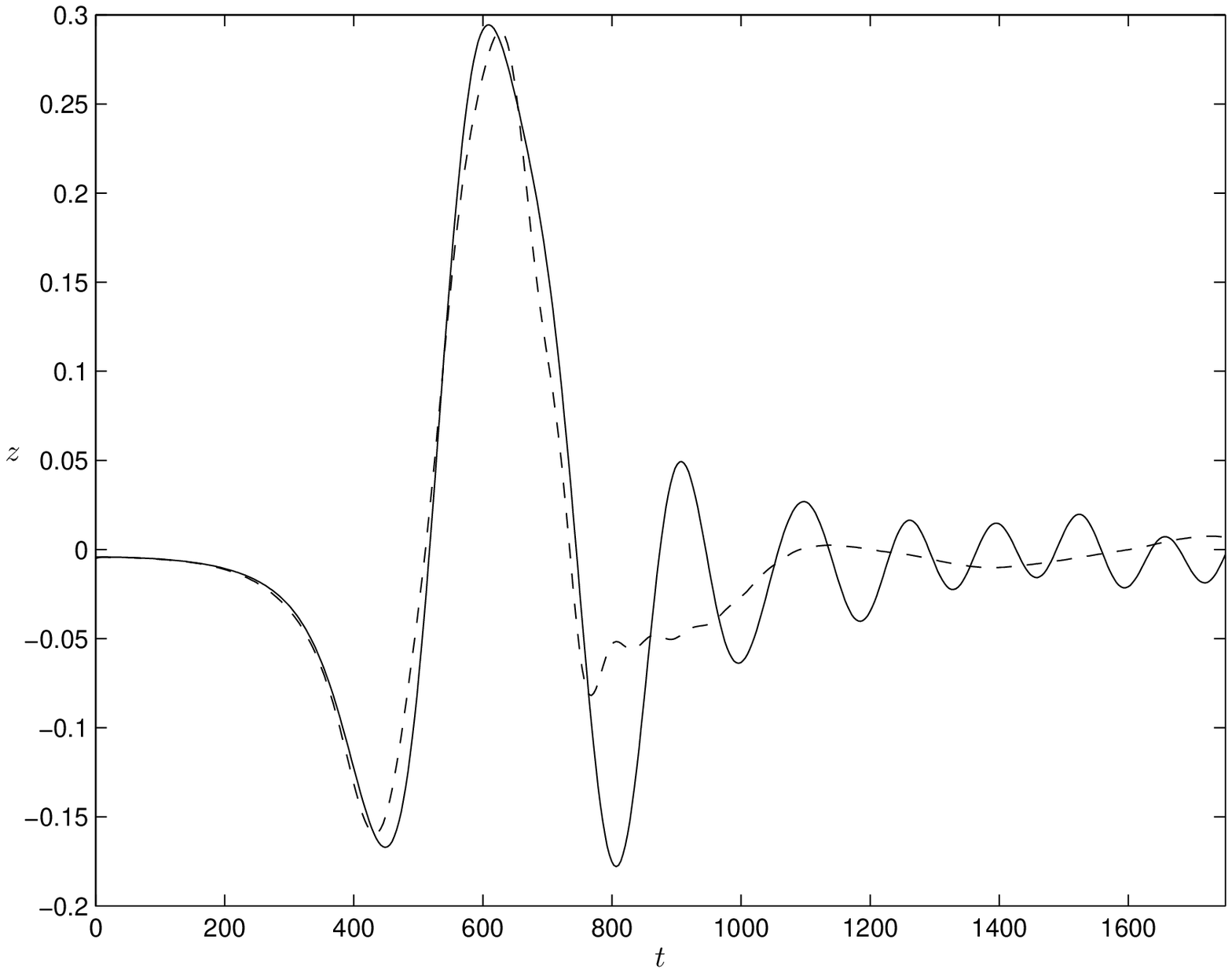}\\
Gauge (iii) & Gauge (iv)\\
\includegraphics[scale=.28,angle=0]{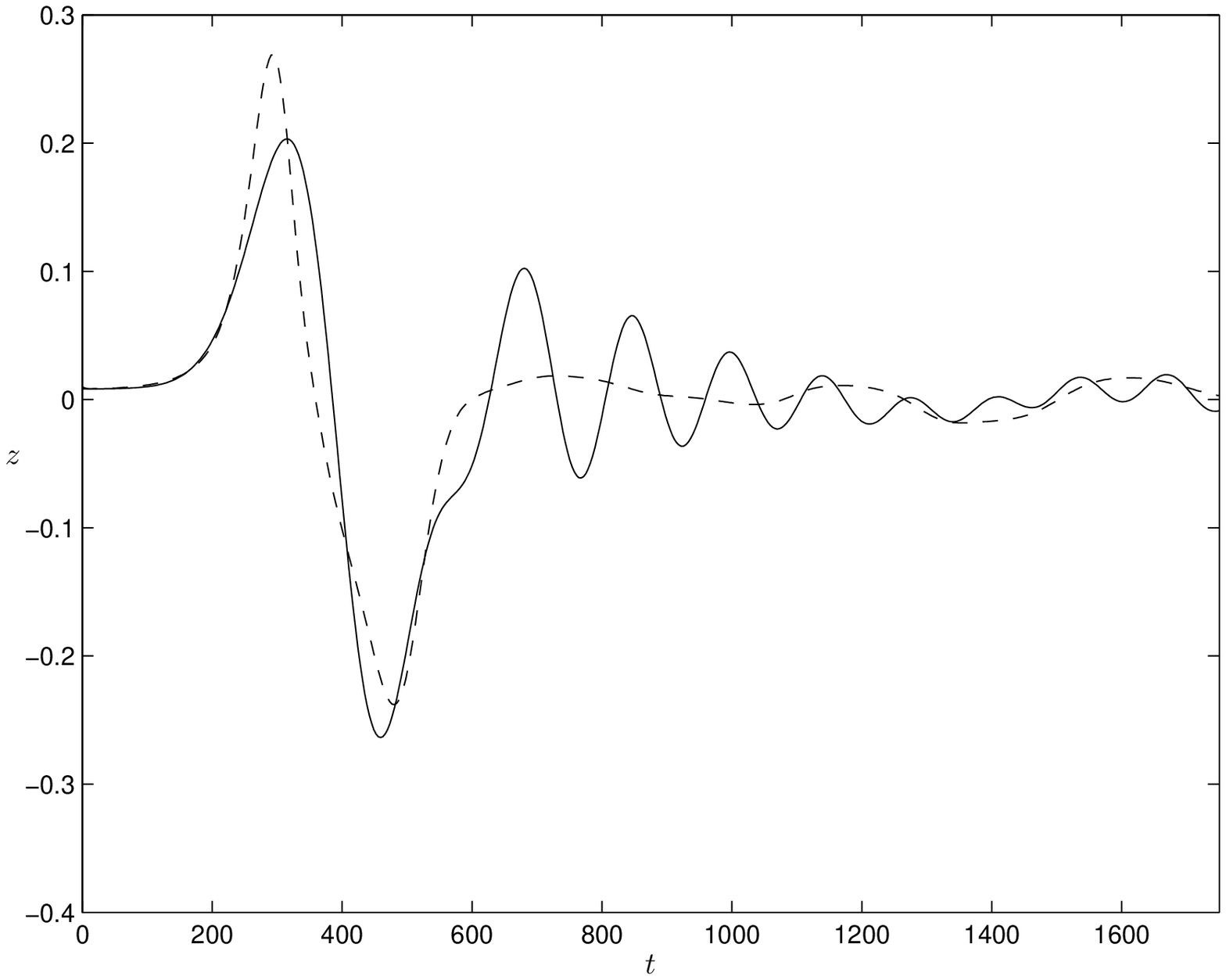}&
\includegraphics[scale=.28,angle=0]{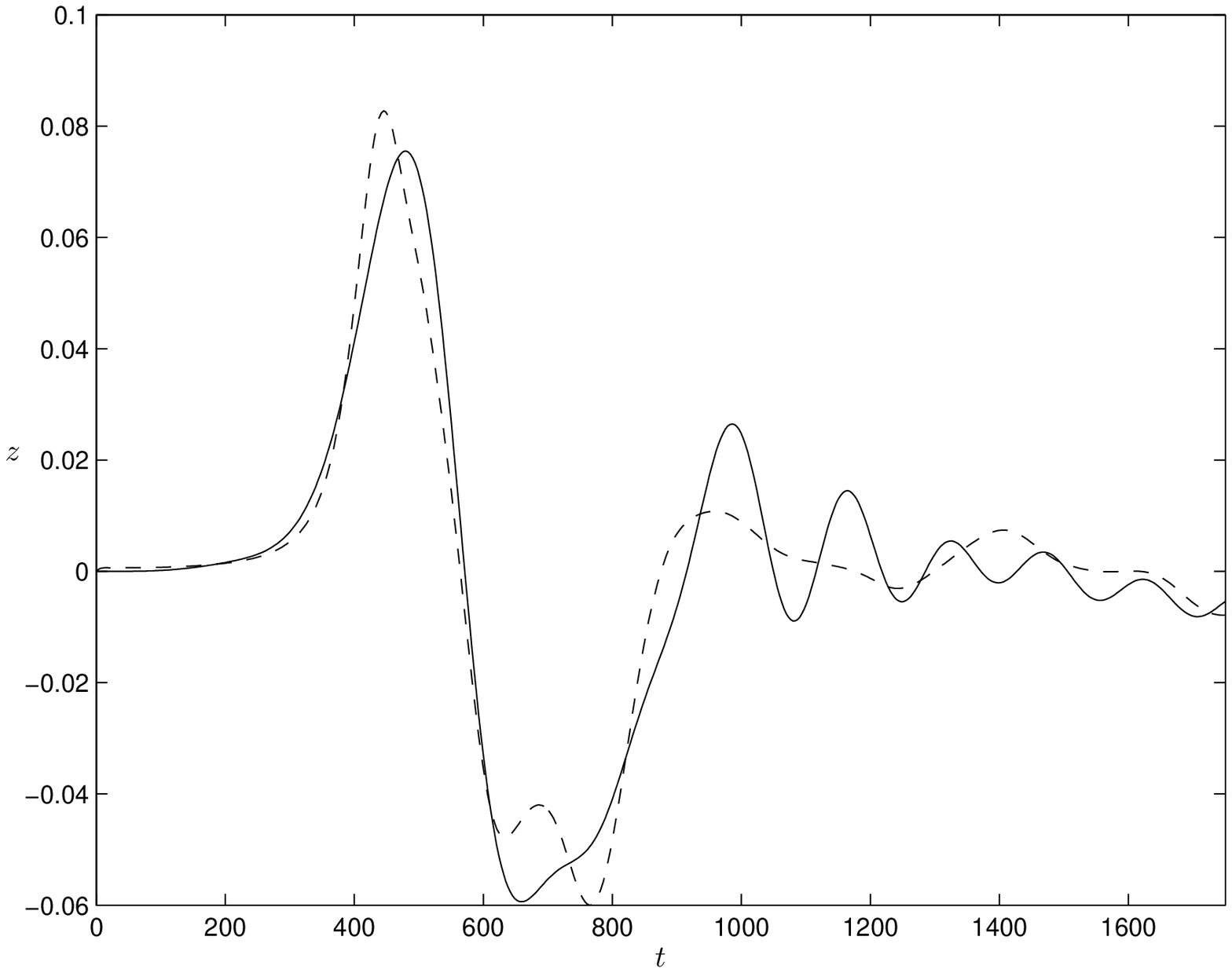}\\
Gauge (v) & Gauge (vi)\\
\includegraphics[scale=.28,angle=0]{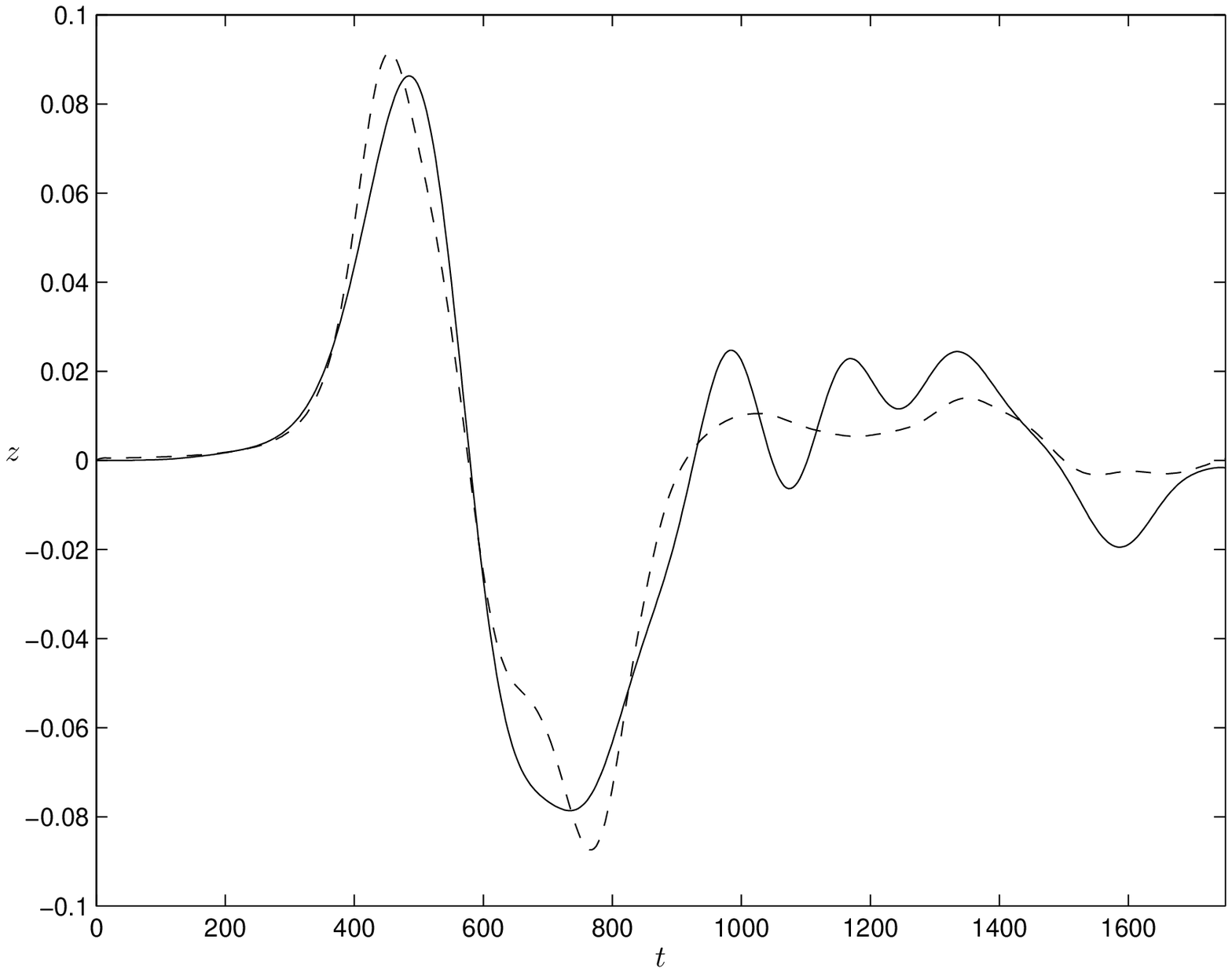}&
\includegraphics[scale=.28,angle=0]{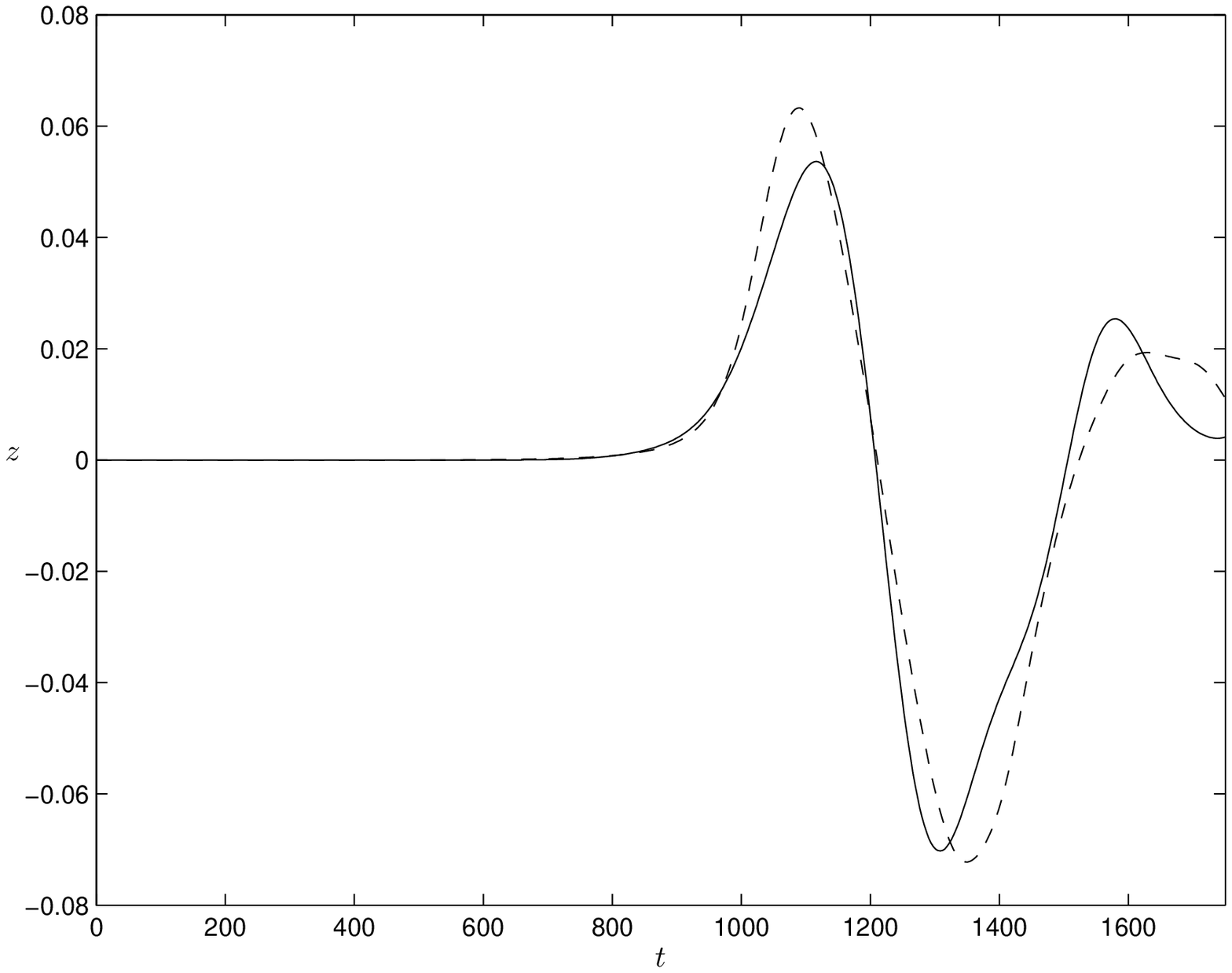}\\
Gauge (vii) & Gauge (viii)\\
\end{tabular}
\end{center}
\caption{$\eta$-component of the solution at the eight wave gauges shown in Figure \ref{F8.1} as a function of $t$. Boussinesq ---, MOST $-\,-$.}\label{F8.3}
\end{figure}

In addition to surface elevation contour plots, we measured the variation of $\eta$ as a function of time at eight wave `gauges' placed at the positions represented by
$\diamondsuit$ in Figure \ref{F8.1}. Specifically, gauges were placed at the points 
(i) $(107.5,-8.5)$, (ii) $(106.5,-8)$, (iii) $(108,-8.5)$, (iv) $(107.3,-8.8)$, (v) $(106.9,-10.6)$, (vi) $(107.7,-11)$, (vii) $(105.9,-10.35)$, (viii) $(104.9,-10.8)$.
The results are shown in Figure \ref{F8.3}.  

\begin{sidewaysfigure}
\begin{center}
\begin{tabular}{cccccc}
\hline
& &  Boussinesq model & & &\\
\includegraphics[scale=.18,angle=0]{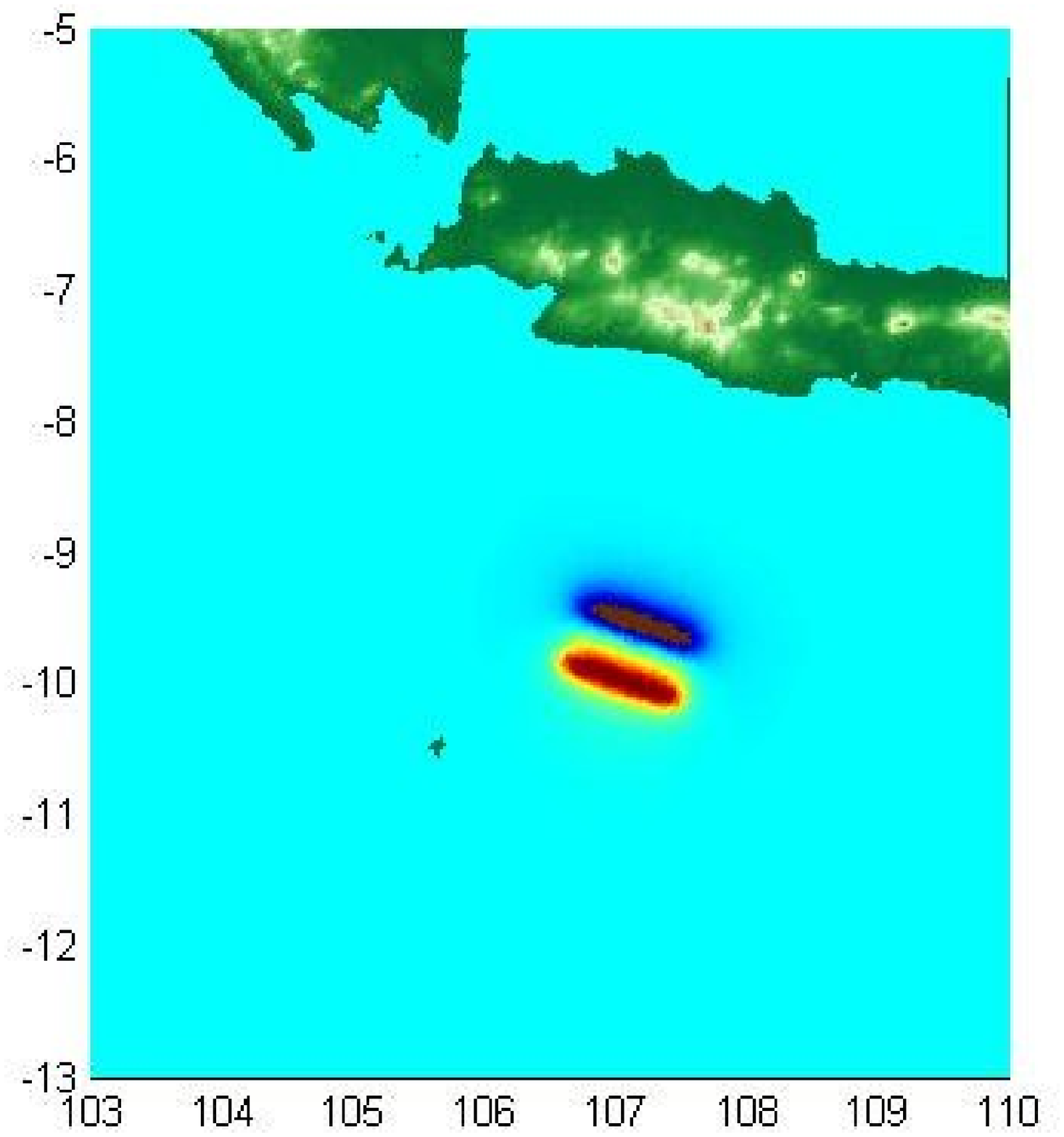}&
\includegraphics[scale=.18,angle=0]{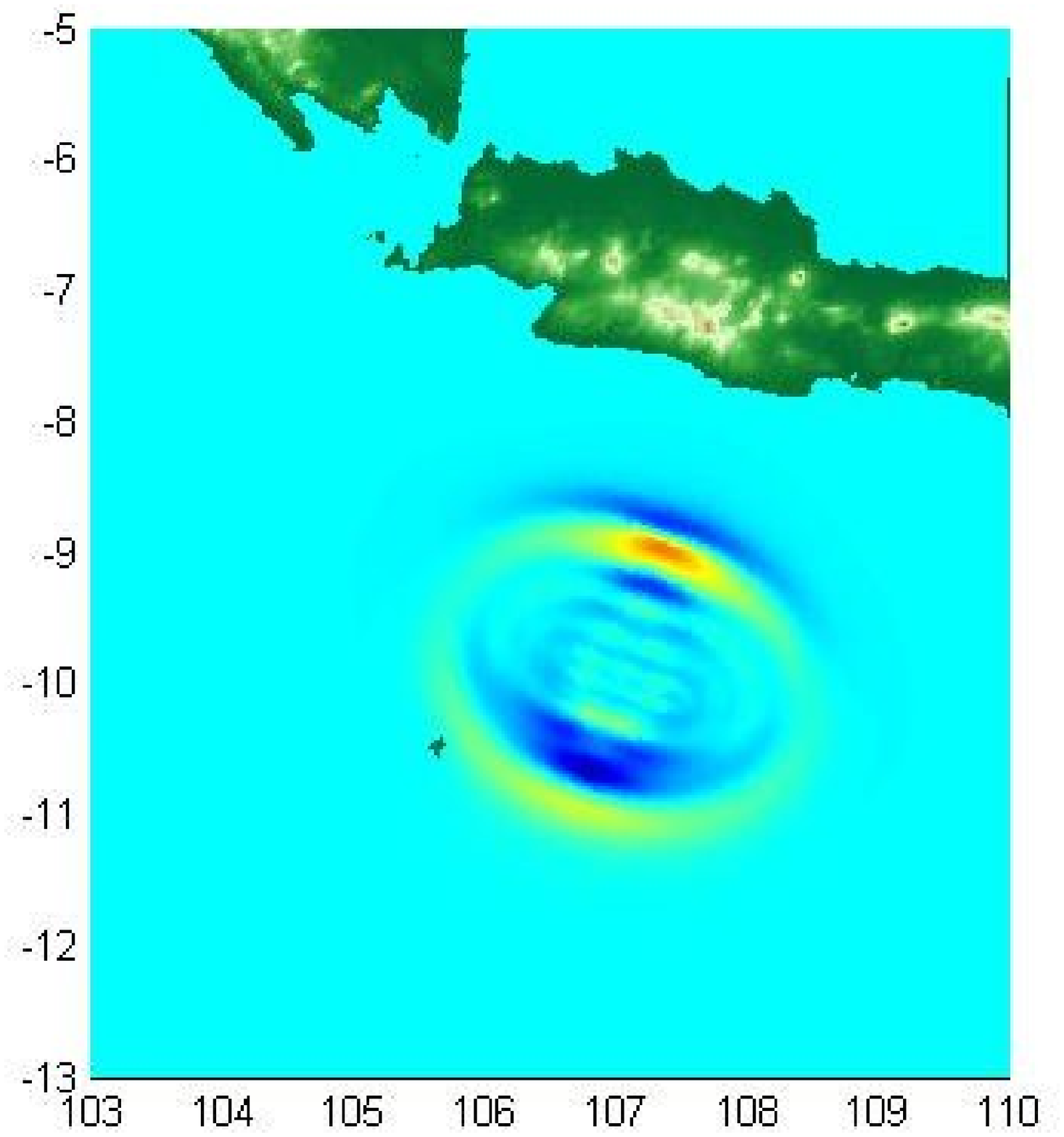}&
\includegraphics[scale=.18,angle=0]{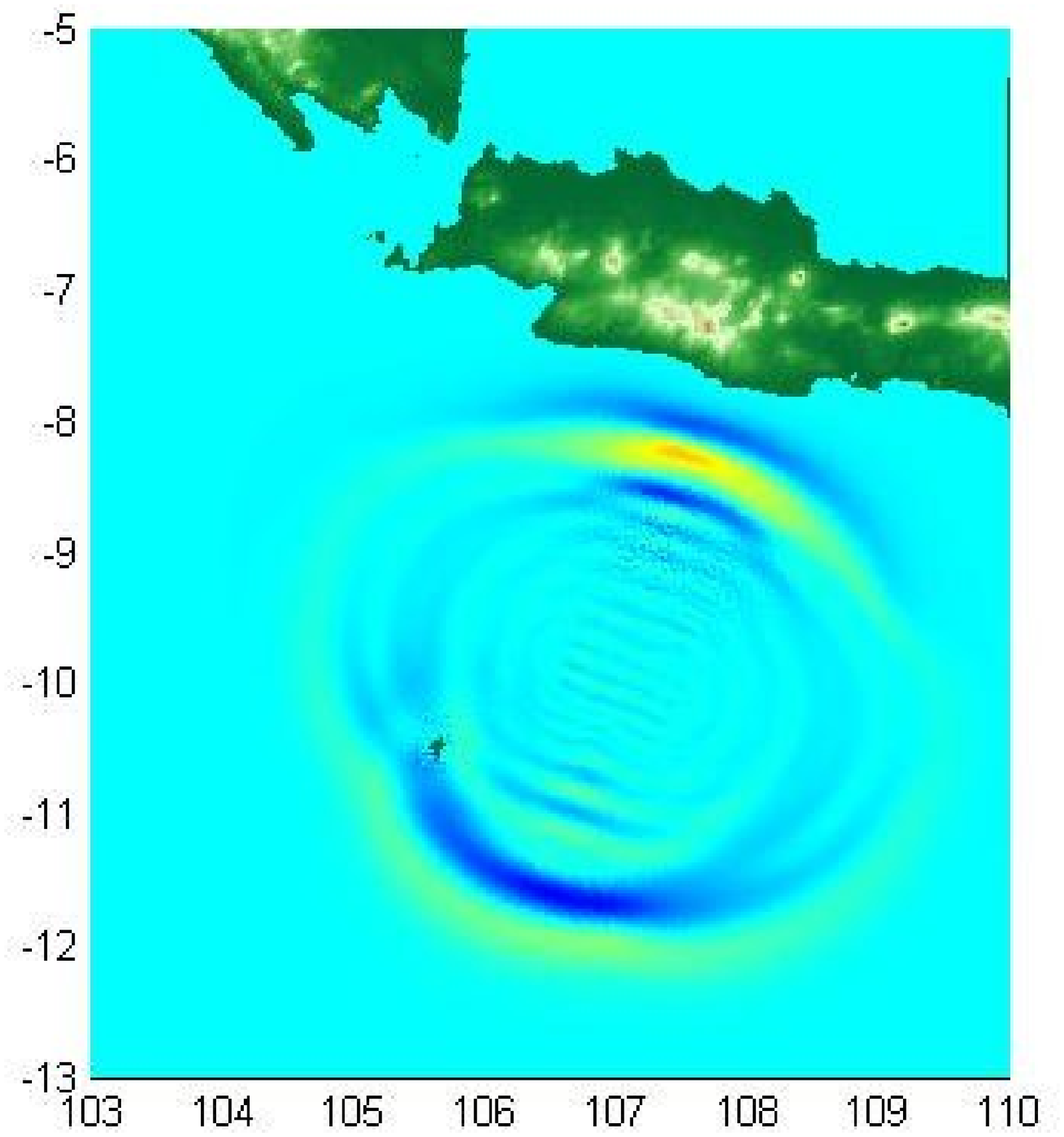}&
\includegraphics[scale=.18,angle=0]{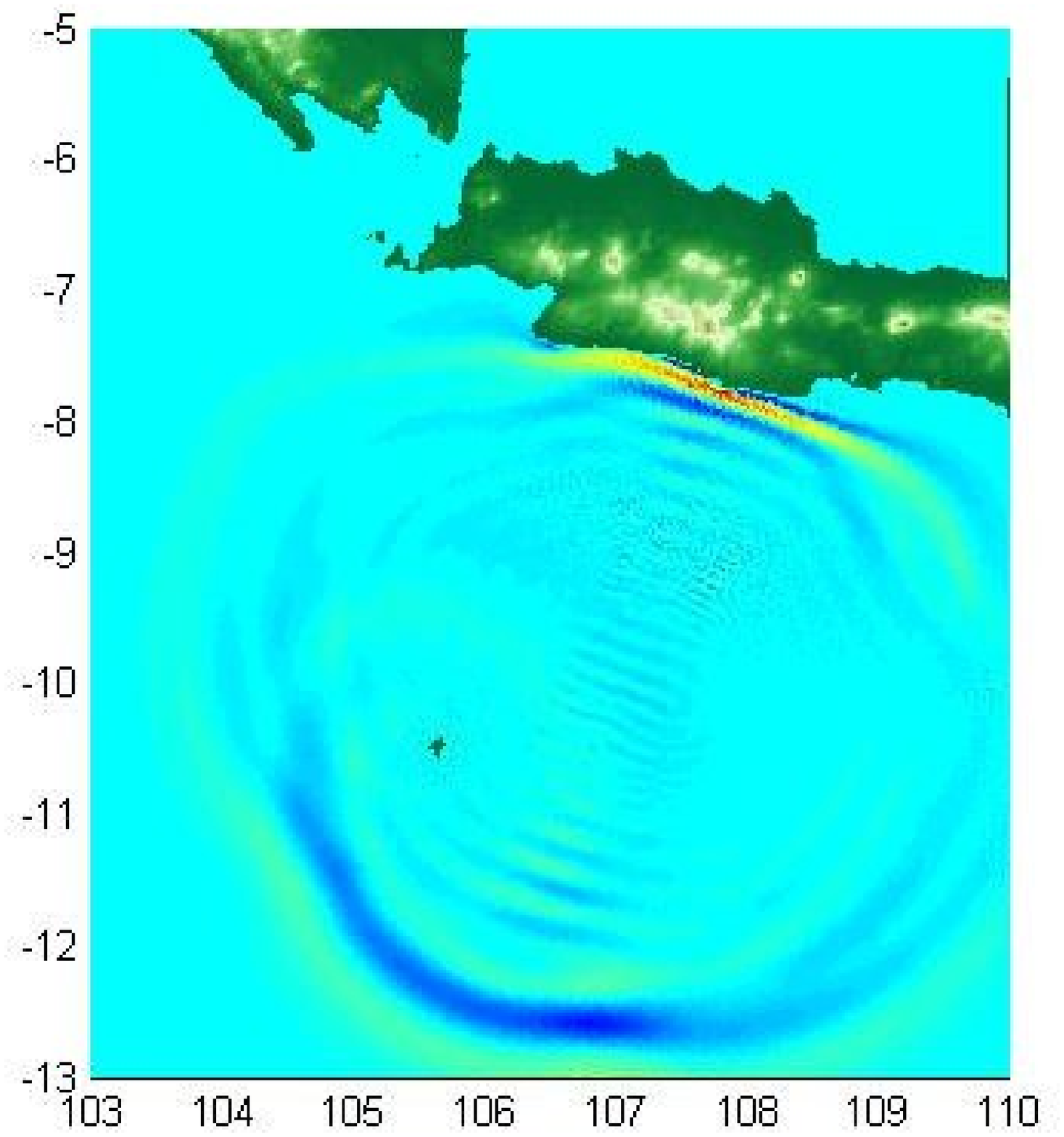}&
\includegraphics[scale=.233,angle=0]{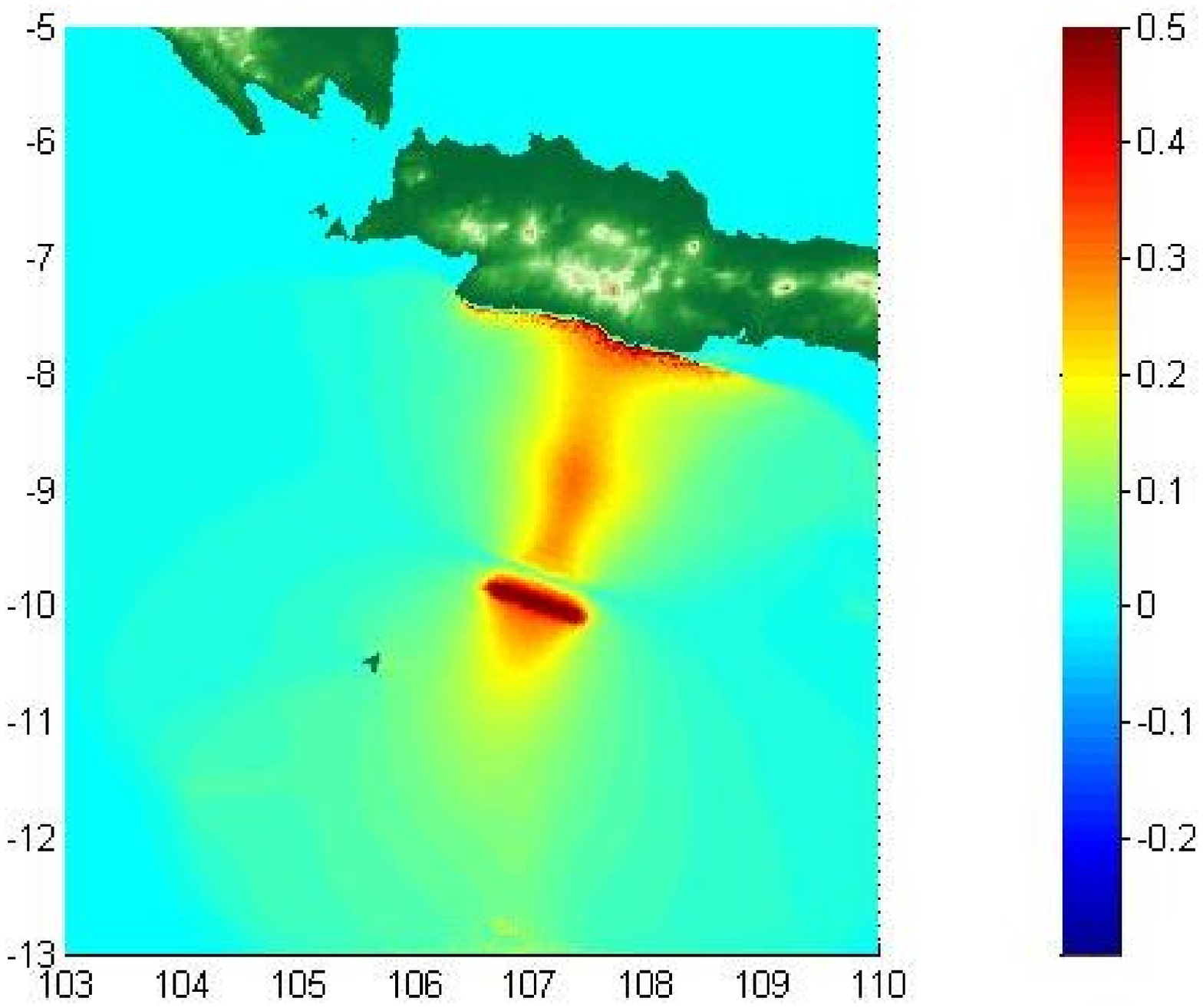}\\
$t=0$sec & $t=500$sec & $t=1000$sec & $t=1500$sec & maximum up to $t=1750$sec\\
\hline
& & MOST model & & &\\
\includegraphics[scale=.18,angle=0]{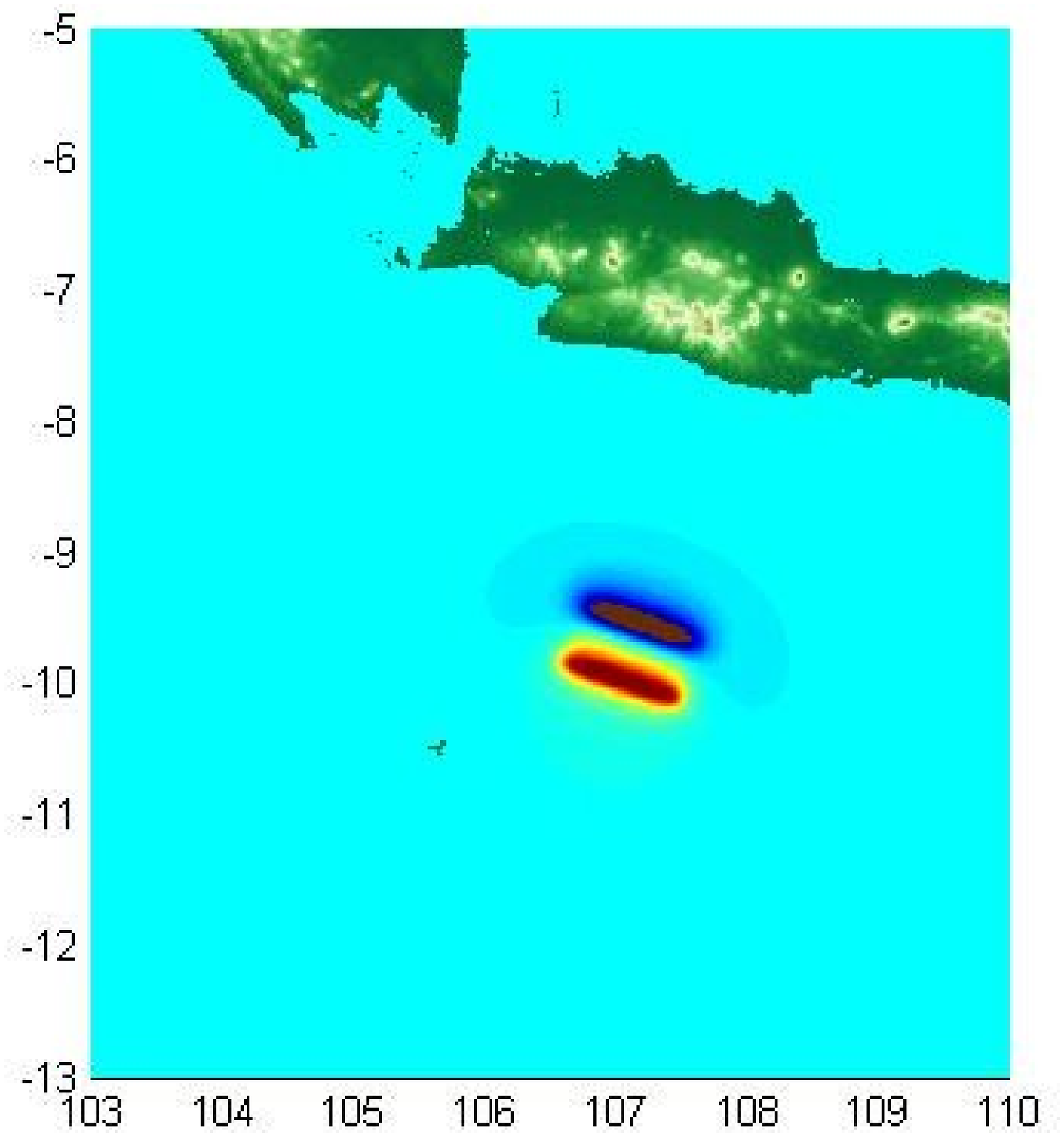}&
\includegraphics[scale=.18,angle=0]{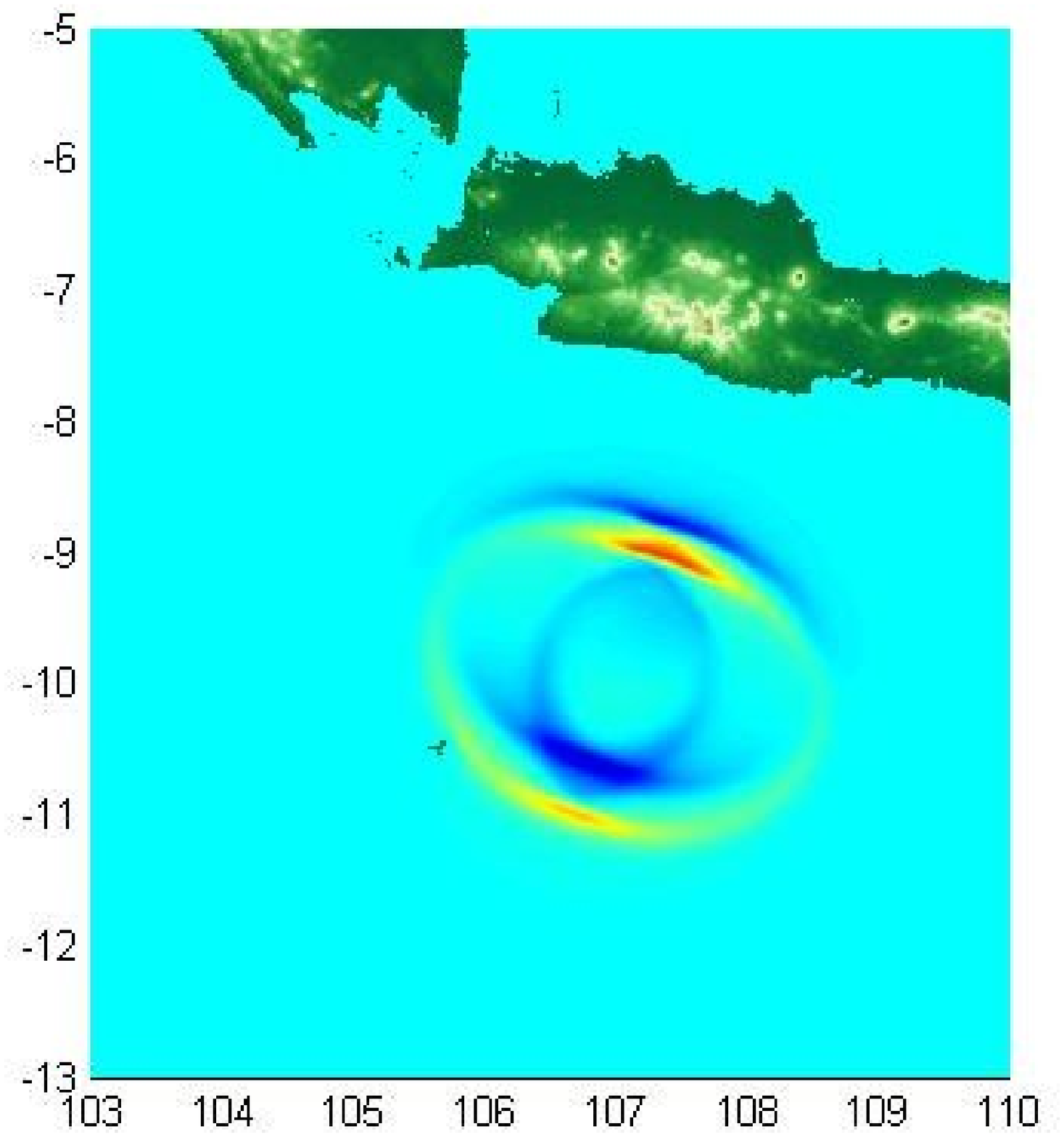}&
\includegraphics[scale=.18,angle=0]{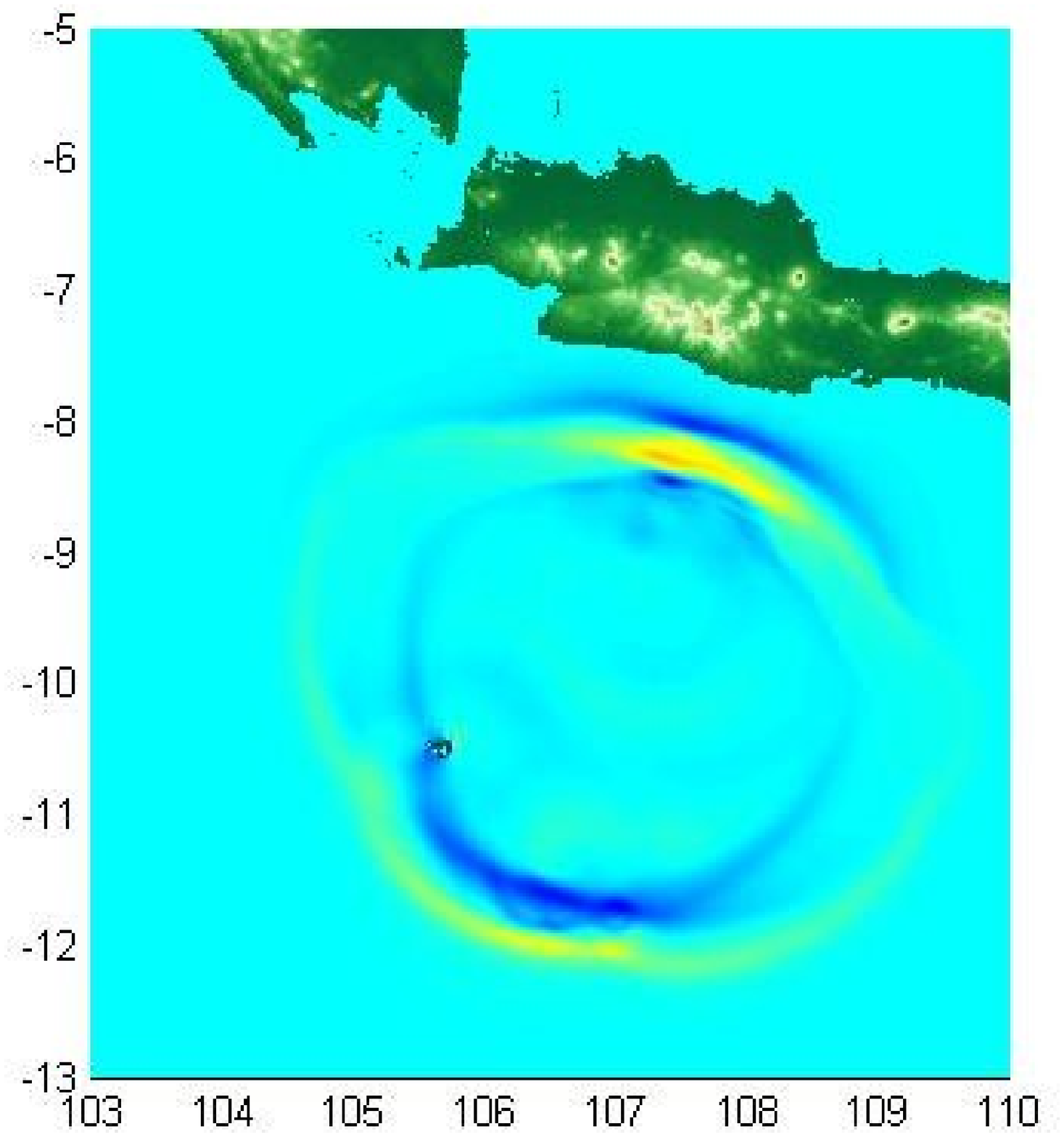}&
\includegraphics[scale=.18,angle=0]{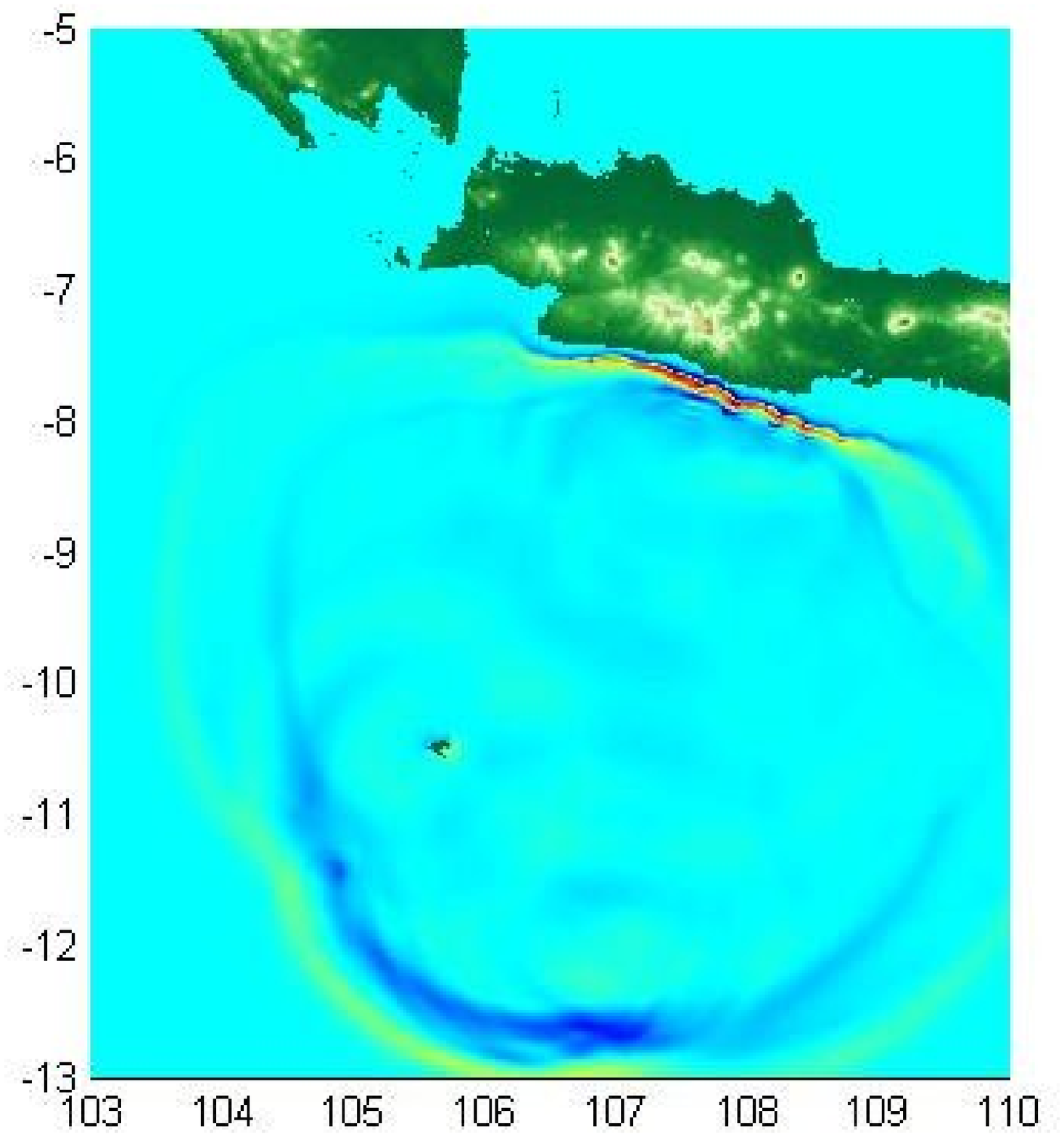}&
\includegraphics[scale=.233,angle=0]{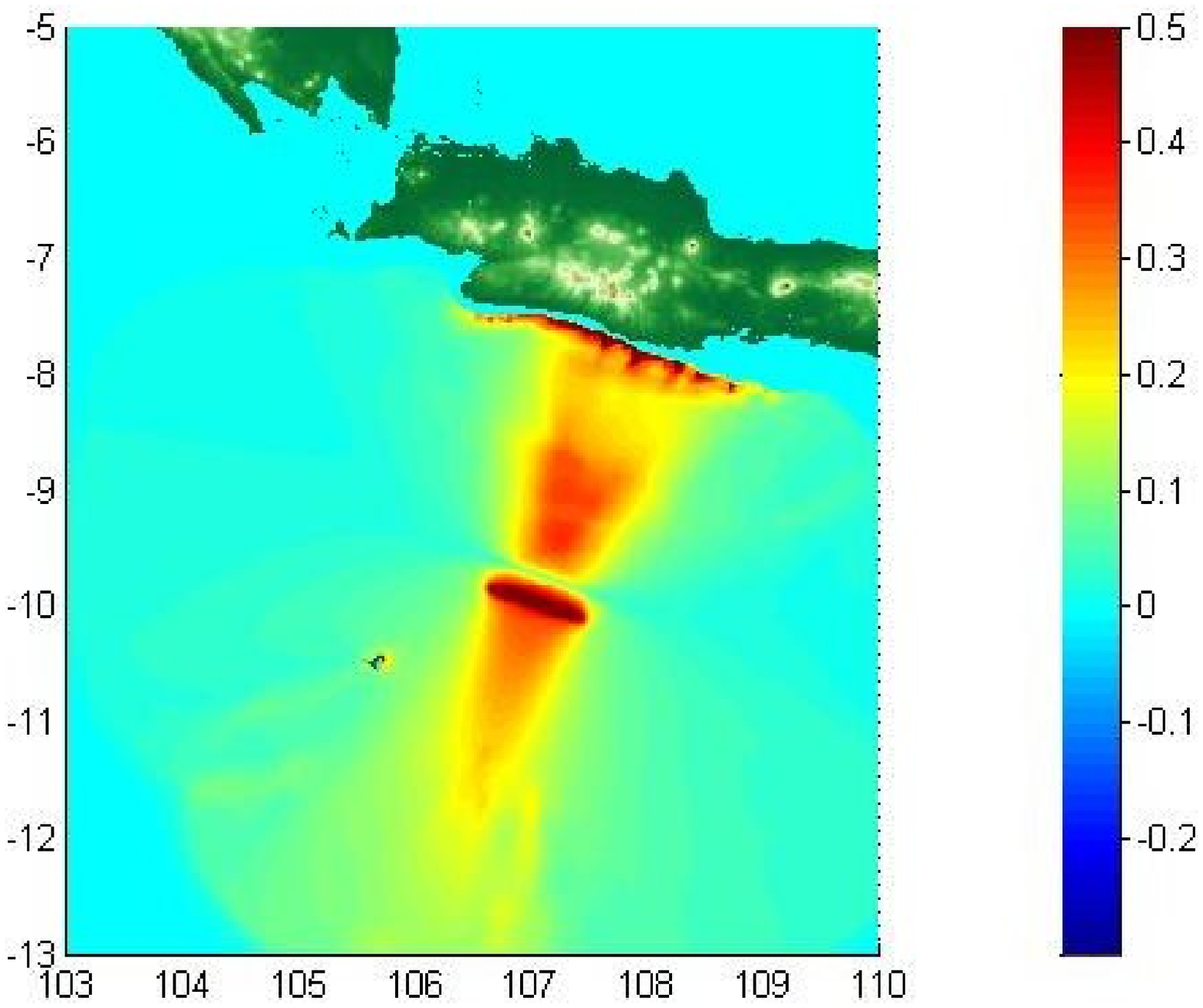}\\
$t=0$sec & $t=500$sec & $t=1000$sec & $t=1500$sec & maximum up to $t=1750$sec\\
\hline
\end{tabular}
\end{center}
\caption{$\eta$ at $t=0,500,1000,1500$sec, and maximum of $\eta$ up to $t=1750$sec. (All distances in degrees. Wave heights in meters).}\label{F8.2}
\end{sidewaysfigure}

In Figure \ref{F8.2} we present the evolution of the initial data and the propagation of the ensuing tsunami in a series of surface elevation contour plots.
We observe that the
tsunami waves produced by the Boussinesq code and MOST model have similar shapes except in the dispersive tail (see also Figure \ref{F8.3}). The rightmost images of Figure \ref{F8.2} represent the maximum of $\eta$ at $(x,y)$ up to $t=1750$sec. (The color scale in the last two graphs of Figure \ref{F8.2} is between the values $-0.3$ and $0.5$ but we checked that the solution, especially near the coastline, exceeds the value of 0.9m.) From these graphs one may observe that in the case of the Boussinesq model the front of the tsunami wave is more narrowly directed than the front produced by MOST. 
In Figure \ref{F8.3} we observe that the amplitudes of the solution of the Boussinesq and the MOST models agree 
quite well at all the gauges. 
We conclude that in this experiment the results of propagation of the tsunami wave produced by the MOST model and by the Boussinesq model are quite similar. For more information about the specific event we refer to \cite{FKM}.

\section*{Acknowledgment}

The author would like to thank Prof. C. Synolakis for valuable advice on tsunamis and MOST. In addition he would like to thank Profs. F. Dias, V. Dougalis, J.-C. Saut, Drs. D. Dutykh, D. Mitsoudis, and Ms. E. Flouri for valuable discussions and comments.

\bibliographystyle{elsart-num-sort}	% (uses file "plain.bst")
\bibliography{myref}		% expects file "myref.bib"

\begin{thebibliography}{10}
\expandafter\ifx\csname url\endcsname\relax
  \def\url#1{\texttt{#1}}\fi
\expandafter\ifx\csname urlprefix\endcsname\relax\def\urlprefix{URL }\fi

\bibitem{Bona}
J.~L. Bona, {Solitary waves and other phenomena associated with model equations
  for long waves}, Fluid Dynamics Transactions 10 (1980) 77--111.

\bibitem{BC}
J.~L. Bona, M.~Chen, {A Boussinesq system for two-way propagation of nonlinear
  dispersive waves}, Physica D 116 (1998) 191--224.

\bibitem{BCS1}
J.~L. Bona, M.~Chen, J.-C. Saut, {Boussinesq equations and other systems for
  small-amplitude long waves in nonlinear dispersive media: I. Derivation and
  Linear Theory}, J. Nonlinear Sci. 12 (2002) 283--318.

\bibitem{BCS2}
J.~L. Bona, M.~Chen, J.-C. Saut, {Boussinesq equations and other systems for
  small-amplitude long waves in nonlinear dispersive media: II. The nonlinear
  theory}, Nonlinearity 17 (2004) 925--952.

\bibitem{BCL}
J.~L. Bona, T.~Colin, D.~Lannes, {Long wave approximations for water waves},
  Arch. Rational Mech. Anal. 178 (2005) 373--410.

\bibitem{Chazel}
F.~Chazel, {Influence of bottom topography on long water waves}, Math. Model.
  Num. Anal. 41 (2007) 771--799.

\bibitem{C3}
M.~Chen, {Equations for bi-directional waves over an uneven bottom}, Math.
  Comput. Simulation 62 (2003) 3--9.

\bibitem{C2}
M.~Chen, {Numerical investigation of a two-dimensional Boussinesq system},
  Discrete Contin. Dynam. Systems 23 (2009) 1169--1190.

\bibitem{DMreview}
V.~A. Dougalis, D.~E. Mitsotakis, {Theory and numerical analysis of Boussinesq
  systems: A review}, in: N.~A. Kampanis, V.~A. Dougalis, J.~A. Ekaterinaris
  (eds.), Effective Computational Methods in Wave Propagation, CRC {P}ress,
  2008, pp. 63--110.

\bibitem{DMS1}
V.~A. Dougalis, D.~E. Mitsotakis, J.-C. Saut, {On some Boussinesq systems in
  two space dimensions: Theory and numerical analysis}, ESAIM, Math. Model.
  Num. Anal. 41 (2007) 825--854.

\bibitem{DMS2}
V.~A. Dougalis, D.~E. Mitsotakis, J.-C. Saut, {On initial-boundary value
  problems for a Boussinesq system of BBM-BBM type in a plane domain}, Discrete
  Contin. Dynam. Systems 23 (2009) 1191--1204.

\bibitem{DD2}
D.~Dutykh, F.~Dias, {Tsunami generation by dynamic displacement of sea bed due
  to dip-slip faulting}, this issue.

\bibitem{DD}
D.~Dutykh, F.~Dias, {Dissipative Boussinesq equations}, C. R. Mecanique 335
  (2007) 559--583.

\bibitem{DD1}
D.~Dutykh, F.~Dias, {Water waves generated by a moving bottom}, in: {Kundu
  Anjan} (ed.), Tsunami and Nonlinear waves, Springer, 2007, pp. 65--95.

\bibitem{FKM}
H.~M. Fritz, W.~Kongko, A.~Moore, B.~McAdoo, J.~Goff, C.~Harbitz, B.~Uslu,
  N.~Kalligeris, D.~Suteja, K.~Kalsum, V.~Titov, A.~Gusman, H.~Latief,
  E.~Santoso, S.~Sujoko, D.~Djulkarnaen, H.~Sunendar, C.~Synolakis, {Extreme
  run-up from the 17 July 2006 Java tsunami}, Geophys. Res. Lett. 34 (2007)
  1--25.

\bibitem{GHM}
M.~Guesmia, P.~Heinrich, C.~Mariotti, {Numerical simulation of the 1969
  Portuguese tsunami by a finite element method}, Natural Hazards 17 (1998)
  31--46.

\bibitem{Ham1}
J.~L. Hammack, {A note on tsunamis: their generation and propagation in an
  ocean of uniform depth}, J. Fluid. Mech. 60 (1973) 769--799.

\bibitem{KDD}
Y.~Kervella, D.~Dutykh, F.~Dias, {Comparison between three-dimensional linear
  and nonlinear tsunami generation models}, Theor. Comput. Fluid Dyn. 21 (2007)
  245--269.

\bibitem{LBLS}
P.~J. Lynett, J.~C. Borrero, P.~L.-F. Liu, C.~E. Synolakis, {Field survey and
  numerical simulations: A review of the 1998 Papua New Guinea tsunami}, Pure
  Appl. Geophys. 160 (2003) 2119--2146.

\bibitem{MMS}
P.~A. Madsen, R.~Murray, O.~R. S{\o}rensen, {A new form of the Boussinesq
  equations with improved linear dispersion characteristics}, Coastal Eng. 15
  (1991) 371--388.

\bibitem{MS}
P.~A. Madsen, O.~R. S{\o}rensen, {A new form of the Boussinesq equations with
  improved linear dispersion characteristics. Part 2: A slow-varying
  bathymetry}, Coastal Eng. 18 (1992) 183--204.

\bibitem{N}
O.~Nwogu, {Alternative form of Boussinesq equations for nearshore wave
  propagationy}, J. Waterway, Port, Coastal, and Ocean Eng. 119 (1993)
  618--638.

\bibitem{Ok1}
Y.~Okada, {Surface deformation due to shear and tensile faults in a half
  space}, Bull. Seism. Soc. Am. 75 (1985) 1135--1154.

\bibitem{Ok2}
Y.~Okada, {Internal deformation due to shear and tensile faults in a
  half-space}, Bull. Seism. Soc. Am. 82 (1992) 1018--1040.

\bibitem{P}
D.~H. Peregrine, {Long waves on a beach}, J. Fluid Mech. 27 (1967) 815--827.

\bibitem{Saad}
Y.~Saad, {Iterative methods for sparse linear systems}, PWS Publishing, Boston,
  MA, 1996.

\bibitem{SM}
H.~A. Sch\"{a}ffer, P.~A. Madsen, {Further enhancements of Boussinesq-type
  equations}, Coastal Engineering 26 (1995) 1--14.

\bibitem{Triangle}
J.~R. Shewchuk, {Triangle: engineering a 2D quality mesh generator and Delaunay
  triangulator}, in: L.~C. Ming, D.~Manocha (eds.), Applied Computational
  Geometry, Springer, New York, 1996, pp. 203--222.

\bibitem{Tphd}
V.~V. Titov, Numerical modeling of long waves, {Ph.D.} {T}hesis, University of
  Southern California (1997).

\bibitem{TS}
V.~V. Titov, C.~E. Synolakis, {Modeling of breaking and non-breaking long-wave
  evolution and run-up using VTCS-2}, J. Waterway, Port, Coastal, Ocean Eng.
  ASCE 121 (1995) 308--316.

\bibitem{TS2}
V.~V. Titov, C.~E. Synolakis, {Numerical modeling of 3-D long wave runup using
  VTCS-3}, in: P.~Liu, H.~Yeh, C.~Synolakis (eds.), Long Wave Runup Models,
  World Scientific, Singapore, 1996, pp. 242--248.

\bibitem{W}
G.~B. Witham, {Linear and Non-linear Waves}, Wiley, New York, 1974.

\bibitem{Wu}
T.~Y. Wu, {Long waves in ocean and coastal waters}, Proceedings of the American
  Society of Civil Engineers 107 (1981) 501--522.

\end{thebibliography}

\end{document}